# In-plane quasi-single-domain BaTiO$_3$ *via* interfacial symmetry engineering

J. W. Lee[1,9], K. Eom[1,9], T. R. Paudel[2,3], B. Wang[4], H. Lu[2], H. Huyan[5], S. Lindemann[1], S. Ryu[1], H. Lee[1], T. H. Kim[1], Y. Yuan[4], J. A. Zorn[4], S. Lei[4], W. Gao[5], T. Tybell[6], V. Gopalan[4], X. Pan[5,7,8], A. Gruverman[2], L. Q. Chen[4], E. Y. Tsymbal[2], and C. B. Eom[1*]

**Abstract**

**The control of the in-plane domain evolution in ferroelectric thin films is not only critical to understanding ferroelectric phenomena but also to enabling functional device fabrication. However, in-plane polarized ferroelectric thin films typically exhibit complicated multi-domain states, not desirable for optoelectronic device performance. Here we report a strategy combining interfacial symmetry engineering and anisotropic strain to design single-domain, in-plane polarized ferroelectric BaTiO$_3$ thin films. Theoretical calculations predict the key role of the BaTiO$_3$ / PrScO$_3$ $(110)_O$ substrate interfacial environment, where anisotropic strain, monoclinic distortions, and interfacial electrostatic potential stabilize a single-variant spontaneous polarization. A combination of scanning transmission electron microscopy, piezoresponse force microscopy, ferroelectric hysteresis loop measurements, and second harmonic generation measurements directly reveals the stabilization of the in-plane quasi-single-domain polarization state. This work offers design principles for engineering in-plane domains of ferroelectric oxide thin films, which is a prerequisite for high performance optoelectronic devices.**

[1]Department of Materials Science and Engineering, University of Wisconsin-Madison, Madison, Wisconsin 53706, USA; [2]Department of Physics and Astronomy & Nebraska Center for Materials and Nanoscience, University of Nebraska, Lincoln, Nebraska 68588, USA; [3]Department of Physics, South Dakota School of Mines and Technology, Rapid City, South Dakota 57701, USA; [4]Department of Materials Science and Engineering, The Pennsylvania State University, University Park, Pennsylvania 16802, USA; [5]Department of Materials Science and Engineering, University of California, Irvine, California 92697, USA; [6]Department of Electronic Systems, Norwegian University of Science and Technology, 7491 Trondheim, Norway; [7]Department of Physics and Astronomy, University of California, Irvine, California 92697, USA; [8]Irvine Materials Research Institute, University of California, Irvine, California 92697, USA; [9]These authors contributed equally to this work.

* Corresponding author. Email: eom@engr.wisc.edu

Oxide interfaces have acquired much attention in the last decade due to the emergence of novel multifunctionalities[1–4]. For example, LaAlO$_3$ and SrTiO$_3$ are both electrical insulators, but when they are grown on top of each other, highly conducting channels are formed at the interface[1], accompanying ferromagnetism[2–5] and gate-tunable superconductivity[6,7]. A key question that arises from such novel interfaces is whether bulk properties of a heterostructure can be controlled through the emergent states of matter at the interfaces. Interfacial effects arising from electrostatic and crystalline symmetry mismatches have been known to play important roles in determining the domain structures of ferroelectric thin films. Indeed, Yu et al. reported that atomically precise control of the interface could result in different polarization states in (001)-oriented ferroelectric films[8], having the polarization perpendicular to the surface. However, mechanisms underlying the interfacial symmetry mismatch on domain states and the ferroelectric response of in-plane polarized films are still not understood.

Single-domain in-plane polarized states are highly desirable for a number of potential functional device applications, such as high-performance electro-optic modulators[9] and planar-type ferroelectric tunnel junctions[10]. However, epitaxial oxide thin films with in-plane polarization typically exhibit complicated multi-domain states which can severely degrade optical device performances[11]. For example, for ferroelectric waveguide applications, the waveguide loss and the electro-optic coefficient of ferroelectric materials are strongly influenced by the optical scattering by domain boundaries that exist in multidomain ferroelectric thin films and the crystalline quality of the films[12–15].

In this work, we report the strategy for in-plane single-domain in BaTiO$_3$ (BTO) ferroelectric thin films via interfacial symmetry and electrostatic potential mismatch using $(110)_O$-oriented PrScO$_3$ (PSO) substrates. Density functional theory calculations and phase-field simulations predict the key roles of the interfacial environment between a film and the substrate, *i.e.*, anisotropic strain, monoclinic distortion, and interfacial electrostatic potential, in stabilizing in-plane single-domain in BTO films on PrScO$_3$ (PSO) $(110)_O$ substrates. Scanning transmission electron microscopy (STEM), piezoresponse force microscopy (PFM), polarization hysteresis loop measurements, and optical second harmonic generation (SHG) results reveal that polarization direction of the BTO film is mainly along PSO $[1\bar{1}0]_O$ direction, while it has a small variable tilting along PSO $\langle 002 \rangle_O$ directions, leading to "quasi-single-domain" state. This work offers a novel approach to engineer in-plane ferroelectric epitaxial oxide thin films, enabling the development of novel device applications.

**Results**

**Modelling of in-plane single-domain BTO**

To design in-plane single-domain ferroelectric film, taking BTO – a canonical ferroelectric material – as a model system, we first discuss universal aspects of epitaxial strain for ferroelectric domain configurations. In general, in-plane polarization of ferroelectric oxide thin films is established under tensile strains[16–19]. However, under an isotropic biaxial strain, such films typically possess complicated multi-domain structures due to the presence of two energetically degenerate ferroelastic variants, resulting in four possible ferroelectric variants (Fig. 1a). Even though a proper anisotropic strain could give rise to single in-plane ferroelastic variant, 180° ferroelectric domains, corresponding to two possible ferroelectric variants, are still energetically equivalent (Fig. 1b).

To energetically stabilize one of the two in-plane polarized states, we note that a symmetry lowering of the BTO film (i.e., monoclinic distortion) via an interfacial symmetry mismatch effect could play a central role (Fig. 1c). It is well known that interfacial oxygen octahedral coupling could initiate structures propagation over extensive distances and subsequently allows to tune the structure of epitaxial film[20]. For example, at the interface between CaTiO$_3$ and NdGaO$_3$, oxygen octahedral coupling breaks the mirror symmetry and stabilize a single domain structure with a single monoclinic tilt direction[20]. Once a monoclinic distortion is introduced near the interface region, an out-of-plane polarization will appear due to the symmetry lowering and the orientation will be determined by the polarity of the substrate surface termination. Such a monoclinic distortion of BTO is likely to gradually relax as the thickness of the film increases [21,22], resulting in a single-domain in-plane ferroelectric state at the region far away from the interface (Fig. 1c).

To test feasibility of this approach, we consider orthorhombic rare-earth scandate (REScO$_3$) substrates (Supplementary Table 1). It should be noted that the pseudocubic (pc) lattice parameter of REScO$_3$ substrates, $a_{pc}$ = 0.394 – 0.405 nm[23,24], is similar to $c$-lattice parameter of bulk tetragonal BTO (4.036 Å)[25]. In addition, $(110)_O$-oriented REScO$_3$ substrates possess a monoclinic nature since the out-of-plane direction ($[110]_O$) is not perpendicular to one of in-plane directions ($[1\bar{1}0]_O$)[26]. Furthermore, a reliable method to obtain complete ScO$_2$-terminated REScO$_3$ has been reported[27]. Among them, $(110)_O$-oriented PSO substrate is a promising candidate for the following two reasons: first, its lattice parameters along $b_{pc}$ (i.e. $[002]_O$) and $c_{pc}$ (i.e. $[1\bar{1}0]_O$) axis are similar to the $b$ (=$a$) and $c$ lattice parameters of bulk tetragonal BTO with misfit strains of +0.38% and −0.25%, respectively (Details in axis notations and the crystallographic relationship between BTO and PSO are represented in Supplementary Figure 1). Secondly, PSO has a high in-plane anisotropic ratio ($c_{pc}/b_{pc}$) of 1.005, which is the largest among all scandates (Supplementary Table 1).

To evaluate the possibility of stabilizing an in-plane single-domain state in BTO/PSO heterostructures, we performed first-principles calculations. The BTO/PSO heterostructure was modeled using BTO lying on the top of ScO$_2$-terminated PSO $[110]_O$ (Fig. 2). Notably, we found that there is a monoclinic distortion in the BTO, and its tilt direction is opposite to that of PSO. Other possible structural configurations of BTO, where the tilt direction is the same as PSO was also calculated (Supplementary Figure 2a). Interestingly, BTO/PSO heterostructures with opposite tilt direction turns out to be energetically favorable by 18.6 mJ/m$^2$ (For more details, see the Supplementary Figure 2 and Note 1). As expected, the out-of-plane polarization component of BTO is pointed "downward" to the interface due to negative charges associated with the ScO$_2$ terminated PSO substrate, while the in-plane polarization component is preferably along PSO$[1\bar{1}0]_O$ direction, resulting in an overall polarization along the diagonal direction. On the other hand, to predict possible ferroelectric domain structures far away from the interface of the BTO thin films, we employed the phase-field method[28]. As suggested by the first-principles calculation, we assume that the initial state of the film has a uniform, in-plane polarization along PSO$[1\bar{1}0]_O$ direction. The interfacial effect associated with the negative charges is considered by introducing a layered charge distribution at the interface. A more detailed description of the model is given in the Method section. A representative two-dimensional section of the phase-field simulation results demonstrate a quasi-single-domain polarization state in the in-plane direction along PSO$[1\bar{1}0]_O$ as shown in Fig. 2b. There is also a spatial distribution of polarization which gradually rotates across the film thickness. Near the substrate-film interface an out-of-plane (toward the substrate) distorted polarization (Fig. 2d) is observed whereas near the film surface the polarization is entirely in-plane (along PSO $[1\bar{1}0]_O$ direction) (Fig. 2c). The out-of-plane polarization gradually decreases through the film thickness while the in-plane polarization magnitude remains nearly constant. This single domain state with in-plane polarization in the BTO/PSO heterostructures is consistent with our model as shown in Fig. 1c, suggesting the possibility to stabilize such a unidirectional in-plane polar state.

**Experimental demonstration of in-plane single-domain BTO**

For an experimental demonstration of BTO with single-domain in-plane polarization, BTO films with a thickness of 50 nm were grown on atomically smooth PSO $(110)_O$ substrates by pulsed laser deposition (PLD) (see Methods for details). The crystallinity of the BTO films on PSO substrates was inferred from four-circle X-ray diffraction (XRD) measurements with a Cu $K_{\alpha 1}$ source (Supplementary Figure 4). From reciprocal space maps of the BTO film around the PSO $(33\bar{2})_O$ and PSO $(240)_O$ reflections, we concluded that the BTO films are fully coherent with the PSO substrate. The lattice parameters of BTO were determined to be $a_{pc,BTO}$ =4.004 Å, $b_{pc,BTO}$ = 4.007 Å and $c_{pc,BTO}$ = 4.026 Å, using pseudocubic notation. Given the bulk lattice parameters, the BTO film is under a compressive strain along PSO $[1\bar{1}0]_O$ direction while under a tensile strain along PSO $[002]_O$ direction (Supplementary Table 1).

To directly observe the polarization configuration of a BTO film, high-angle annular dark-field (HAADF) imaging using scanning transmission electron microscopy (STEM) was employed. Figure 3a shows low magnification HAADF-STEM images of the BTO film along the PSO $[002]_O$ zone-axis. Ti displacements are represented by arrows in Fig. 3b, c, e, f, where the size of the arrows corresponds to the magnitude of atomic displacement in each unit cell. The average value of the out-of-plane component of the Ti displacement in the middle region of the film is negligibly small (–0.6 ± 0.9 pm), while strong downward polarization near the interface region (–8.3 ± 1.0 pm) is observed (Fig. 3d). The result of a decreasing out-of-plane polarization towards the middle region of the film is consistent to phase field simulation results as shown in Fig. 2b-d. Remarkably, the BTO film has a unidirectional in-plane Ti displacement (Fig. 3g) along PSO $[1\bar{1}0]_O$ direction in the middle region of the film (Fig. 3e) and BTO/PSO interface region (Fig. 3f). The average in-plane Ti-displacement for each region is estimated to be 6.0 ± 1.1 pm and 6.0 ± 1.0 pm, respectively, which is comparable to the Ti displacement along the polarization direction in bulk single crystal BTO[29]. The detailed overall polarization state of the BTO film is described in Supplementary Note 2. Furthermore, to verify the PSO $(110)_O$ substrate termination, atomic-resolution energy dispersive spectroscopy (EDS) elemental mapping was performed together with HAADF-STEM imaging at the BTO/PSO interface. Supplementary Figure 7 clearly shows that the interface configuration is $ScO_2$-BaO. It should be noted that the TEM data distinctly demonstrate that the polarization directions near the interface are downward (Fig. 3c, d), which is originated from the negative polarity of the substrate surface. Consistent results are obtained from the multiple spots of the sample, indicating that the surface of the PSO substrate is uniformly $ScO_2$ terminated. This is not surprising, because after high-temperature annealing of rare-earth scandate substrates for sufficient time, $ScO_2$ is known to be more stable than the corresponding rare-earth oxide at the surface[27,30].

By ferroelectric polarization hysteresis measurements, we explored the as-grown polarization state of the BTO/PSO heterostructures. The schematics of the electrode configurations for in-plane polarization hysteresis measurements are shown in Fig. 4a. The measurements were performed along two directions: PSO $[1\bar{1}0]_O$ and PSO $[002]_O$. Notably, a well-defined hysteresis loop is observed only when applying electric field parallel to PSO $[1\bar{1}0]_O$, clearly showing the existence of ferroelectricity (Fig. 4b). The overall positive slope of the polarization in the field region from -30 to 30 kV cm$^{-1}$ is attributed to the PSO substrate (Supplementary Figure 8). The hysteresis loops are off-centered toward the negative field direction, pointing to a large polarization imprint[31] mainly due to the asymmetric energy barrier between two polarization directions, *i.e.*, PSO $[1\bar{1}0]_O$ and PSO $[\bar{1}10]_O$. The polarization signal obtained with electric field parallel to PSO $[002]_O$ direction reveals that there is no remanent ferroelectric polarization along this direction (Fig. 4c). However, the non-linear and weak hysteresis behavior of the P-E loops imply a possible polarization rotation along the PSO $\langle 002 \rangle_O$ direction. The origin of this polarization rotation will be discussed later. From the STEM results, we found that the polarization direction of as-grown BTO is along

PSO $[1\bar{1}0]_O$. This is in good agreement with the observation that there is no hysteresis when we apply a positive electric field toward PSO $[1\bar{1}0]_O$ since the polarization of BTO is already saturated in this direction in the as-grown state. Only under negative electric field a hysteresis loop is formed, indicating polarization reversal towards PSO $[\bar{1}10]_O$. It should be noted that when the field sweep is completed, the BTO film has its original polarization state, that is, a polarization towards PSO $[1\bar{1}0]_O$ is still retained. This is a strong evidence for the in-plane polarization state toward PSO $[1\bar{1}0]_O$ in our BTO samples is driven by the interfacial electrostatic potential which is not altered under bias conditions.

To better understand the microscopic domain structure, we performed angular-dependent lateral PFM measurements when the BTO film sample was rotated by a discreet angle with the respect to the surface normal (Fig. 5). As-grown BTO does not show any contrast in vertical PFM phase image mode (Supplementary Figure 9). However, angular dependent lateral PFM measurements show that there are measurable domain features depending on the relative sample-cantilever orientation, where the lateral PFM signal is sensitive to the projected polarization along the axis perpendicular to the cantilever arm. Detailed analysis of the PFM images (Supplementary Figure 10) reveals that the polarization is mainly along PSO $[1\bar{1}0]_O$, consistent with the STEM results (Fig. 3g) and the polarization hysteresis measurements (Fig. 4a). However, there are small periodical zigzag-type changes in the polarization orientation (typically less than 15°) with respect to the PSO $[002]_O$ and $[00\bar{2}]_O$ directions resulting in a quasi-stripe domain structure with a lateral periodicity in the order of 50 nm and overall in-plane domain pattern (Fig. 5). The detailed analysis and interpretation on the PFM imaging of domain structure along PSO $\langle 002 \rangle_O$ are described in Supplementary Note 3.

**Discussion**

The presented data highlights that in-plane polarized single-domain BTO thin film is achievable by control of anisotropic strain, monoclinic distortion due to structural mismatch at an interface, and interfacial valence mismatching. Near the BTO/PSO interface, presence of both in-plane and out-of-plane polarization components result in a diagonal direction of polarization whereas the polarization of BTO in the middle region of the film is entirely in-plane (along PSO $[1\bar{1}0]_O$). The out-of-plane polarization gradually decreases through the film thickness (Fig. 3d), while the in-plane polarization magnitude remains almost constant (Fig. 3g). This is fully consistent with structural analysis showing that the BTO films have slight monoclinic distortions towards the BTO $[00\bar{1}]_{pc}$ direction (Supplementary Figure 17), while there is no tilting along BTO $[0\bar{1}0]_{pc}$ direction (Supplementary Figure 18). Such structural distortions are in accordance with the monoclinic symmetry probed by optical SHG measurements (Supplementary Figure 20–22). From the local electron diffraction patterns, the monoclinic tilting angles between BTO $[001]_{pc}$ and BTO $[100]_{pc}$ near the interface region and film surface region are 91.11° and 90.58°, respectively (Supplementary Figure 17e, f), indicating the structural relaxation far away from the interface, which is also confirmed by geometric phase analysis (GPA) from TEM data (Supplementary Note 4).

We now discuss the polarization rotation along the PSO $\langle 002 \rangle_O$ direction. This is mainly due to the absence of a suitable single crystal substrate which has the same $b$ (=$a$) and $c$ lattice parameters of bulk tetragonal BTO. Tensile strain along the PSO $[002]_O$ as shown in Supplementary Table 1 could give rise to such a non-zero polarization state, leading to the zigzag-type as-grown domain structure (Fig. 5f). The polarization rotation in as-grown state results in non-linear and weak hysteresis behavior in P-E loop with applying electric field along the PSO $[002]_O$ direction (Fig. 4c). The origin of non-zero polarization along the PSO $[002]_O$ direction is further supported by phase field simulation where pure in-plane single-domain state is achieved by simply eliminating tensile strain along the PSO $[002]_O$ direction (Supplementary Figure 13-15). Therefore, we concluded that the strategy presented in this work effectively stabilizes single ferroelectric variant along the PSO $[1\bar{1}0]_O$ direction, even though unavoidable extra variants still exist (Fig. 5g).

The strategy we present here is to utilize the effect of an out-of-plane parameter (electric field) to manipulate the in-plane properties (polarization) by lowering the symmetry. We have experimentally demonstrated this strategy by preparation of BTO thin films on PSO $(110)_O$ substrates where quasi-single-domain state along PSO $[1\bar{1}0]_O$ direction is achieved at room temperature. We anticipate our strategy for reducing undesirable variants also to be generally used for tuning in-plane properties of correlated materials where their emergent states are subject to domain evolutions.

## Methods

### Sample growth and characterization

The BTO films were grown on $(110)_O$-oriented PSO substrates by PLD with in-situ reflection high-energy electron diffraction (RHEED) monitoring (Supplementary Figure 4a). The BTO ceramic target was ablated using a KrF (248 nm) excimer laser at a repetition rate of 3 Hz with the laser fluence of ~2 J/cm$^2$. The substrate temperature was kept at 680 ºC with an oxygen partial pressure of 120 mTorr during the growth. The substrate-to-target distance was 60 mm. In all experiments, $(110)_O$-oriented PSO substrates provided by CrysTec with miscut angles of ~ 0.1° were used. The PSO substrates were soaked in deionized water for 30 min, and then annealed at 1100 ºC in an oxygen atmosphere for 3 h.

### DFT calculations

Density functional theory calculations were performed using the projector augmented plane-wave method and the Perdew-Burke-Ernzerhof exchange-correlation functional[32] as implemented in the Vienna *ab-initio* simulation package (VASP)[33,34]. A bilayer structure was employed in the calculations with a ferroelectric (100) BTO layer lying on top of an orthorhombic (110) PSO layer. The theoretical lattice constants of bulk PSO: $a$ = 5.73 Å, $b$ = 5.75 Å, $c$ = 8.13 Å were used. We fixed atomic coordinate of first three layers of PSO to their bulk values and relaxed the rest consisting of one and half layer of PSO and eight layers of BTO (Fig. S2) with the energy convergence limit of $1\times 10^{-4}$ eV. In the calculations, a kinetic energy cutoff of 340 eV was used for plane-wave expansion and a 4 × 4 × 1 Monkhorst-Pack grid[35] of **k** points was used for Brillouin zone integration.

### Phase-field simulations

In the phase-field model, imposed misfit strain is calculated to be anisotropic by using the pseudocubic lattice constants between BTO and PSO. To consider the absence of electrodes on top and bottom of the film, we adopt the open-circuit boundary condition for the electrostatic equation. Then, the system is relaxed to reach an equilibrium state by evolving the polarization configuration with the time-dependent Ginzburg-Landau equation[28].

The ferroelectric BTO is described by the spatial distribution of the spontaneous polarization $\boldsymbol{P}(x, y, z)$, the relaxational kinetics of which follows the time-dependent Ginzburg-Landau equation,

$$\frac{\delta \boldsymbol{P}}{\delta t} = -L \frac{\delta F}{\delta \boldsymbol{P}}. \tag{1}$$

Here, $F$ is the free energy which consists of elastic ($f_{\text{elastic}}$), electrostatic ($f_{\text{electrostatic}}$), bulk ($f_{\text{bulk}}$), and gradient ($f_{\text{gradient}}$) energy contributions, and $L$ is the kinetic coefficient. The total free energy can be described as,

$$F = \int_V (f_{\text{bulk}} + f_{\text{elastic}} + f_{\text{electrostatic}} + f_{\text{gradient}}) \, dV. \tag{2}$$

More in-depth description of each term in the energy contributions can be found in the references[36]. The parameters of the model for BTO follows from Li *et al.*[37].

The 3D system is discretized into a $128\Delta x \times 128\Delta y \times 90\Delta z$ grid with $\Delta x = \Delta y = \Delta z = 1$ nm. The thickness of the substrate takes up $30\Delta z$ to accommodate the relaxation of the displacement arising from the domain structure of the 50 nm film. The free surface of the film is described by the traction-free boundary condition on the top while the bottom interface is assumed to be fully coherent. As a result, the prescribed misfit strains between the film and the substrate can be calculated by using the pseudocubic lattice constant $a_{\text{BTO}}^{\text{pc}} = (a_{\text{BTO}}^2 c_{\text{BTO}})^{\frac{1}{3}}$ of BTO and that of the (110)$_\text{O}$ plane of PSO, i.e., $\varepsilon_{xx} = \frac{a_{\text{PSO}}^{[\bar{1}10]} - a_{\text{BTO}}^{\text{pc}}}{a_{\text{BTO}}^{\text{pc}}} = 0.50\%$ and $\varepsilon_{yy} = \frac{a_{\text{PSO}}^{[002]} - a_{\text{BTO}}^{\text{pc}}}{a_{\text{BTO}}^{\text{pc}}} = 0.01\%$. As the lattices of the film and the substrate on the interface are orthogonal, there is no in-plane shear misfit strain, i.e., $\varepsilon_{xy} = \varepsilon_{yx} = 0.0\%$. Considering the absence of top and bottom electrodes of the BTO film in experiments, we adopt the open-circuit boundary condition at both top and bottom of the film, i.e., $\boldsymbol{D} \cdot \boldsymbol{n} = 0$, where $D$ is the electric displacement and $\boldsymbol{n}$ is the normal of the interface plane. The elastic and electric equilibrium equations for the thin-film system are solved following the method developed by Li *et al.*[38]. To account for the interfacial effect as predicted by the first-principles calculations, we assume an effective charged layer at the bottom interface with a fixed, uniform charge density $\rho = -0.5$ C/m$^2$. The phase-field simulations were performed by using the phase-field package (mupro.co).

**STEM measurements**

TEM specimens were prepared by mechanical polishing followed by argon ion milling using Gatan PIPS II. STEM HAADF imaging and EDS experiments were carried out on a JEOL Grand ARM300CF equipped with a cold field emission gun and double spherical aberration correctors with a spatial resolution of ~ 0.7 Å operating at 300 keV in Irvine Materials Research Institute at the University of California, Irvine. STEM images were taken with the convergence angle of the incident electrons at 32 mrad and the collection angle at 90–165 mrad. EDS mappings were acquired using dual silicon-drift detectors (SDDs). 50 scans (each with a 0.4 ms dwell time) in the same area across the interface were summed. The high resolution HAADF STEM imaging provides spatial resolution adequate to measure the atomic positions of the A and B site cations of BTO. The high frequency noise was removed by applying an annular mask in frequency space, and then the initial peak positions were determined by identifying local maxima and refined by fitting Gaussian curves to obtain the atom center positions. Displacements were calculated as the difference between the center of each cation and the center of mass of its for adjacent neighbors. Strain analysis is based on GPA[39] that preinstalled in Gatan Digital Micrograph software.

**PFM measurements**

PFM measurements were performed on a commercial AFM system (MFP3D, Asylum Research). Conductive Pt/Ir coated probes (PPP-EFM, Nanosensors) were used for imaging, with an ac modulation amplitude of 0.6 V at a frequency around 350 kHz in the resonance enhanced PFM mode. Both in-plane and out-of-plane signal were collected to examine the possible domain structures.

**P-E loop measurements**

Polarization vs. Electric field (PE) hysteresis loops were measured using a Radiant Technologies Precision Premier II Ferroelectric Testing system with a 4 kV High-Voltage Interface and a 4 kV TREK amplifier. PE loops were performed using a 250–400 Hz frequency, sinusoidal waveform with an amplitude of 150 V. To make contact to the sample electrodes we used micromanipulator probe tips with a diameter of 10 μm.

**SHG measurements**

To probe the symmetry of BTO film, optical second harmonic polarimetry measurement was performed using a far-field setup as schematically shown in Supplementary Figure 20. Linear polarized optical femtosecond pulses at $\lambda$ = 800 nm (100 fs, 1 kHz) with polarization direction $\varphi$ was focused onto the sample at an incident angle $\theta$. The X-polarized ($I_{2\omega,X}$) and Y-polarized ($I_{2\omega,Y}$) components of second harmonic signal was collected by a photo-multiplier tube behind a band pass filter. Polarimetry measurement on BTO film was performed by rotating the incident polarization $\varphi$ at each $\theta$ for two different sample orientations (***O1:*** ($S_1$, $S_2$, $S_3$) = ($[00\bar{1}]_O$, $[1\bar{1}0]_O$, $[\bar{1}\bar{1}0]_O$), ***O2:*** ($S_1$, $S_2$, $S_3$) = ($[1\bar{1}0]_O$, $[001]_O$, $[\bar{1}\bar{1}0]_O$)). During measurement, incident angles at $\theta$=-45°, -30°, -15°, 0°, 15°, 30° and 45° were used.

Symmetry analysis of the SHG polarimetry was performed using an analytical model as described in literatures[40]. Theoretical fits to all SHG polarimetry data were simultaneously performed by assuming a single ferroelastic domain with monoclinic symmetry of the BTO film, where the SHG coefficients can be written as:

$$d^m = \begin{pmatrix} 0 & 0 & 0 & 0 & d_{15} & d_{16} \\ d_{21} & d_{22} & d_{23} & d_{24} & 0 & 0 \\ d_{31} & d_{32} & d_{33} & d_{34} & 0 & 0 \end{pmatrix} \qquad (3)$$

**Data availability**

The data that support the findings of this study are available from the corresponding author on reasonable request.

## Acknowledgments

This research is funded by the Gordon and Betty Moore Foundation's EPiQS Initiative, Grant GBMF9065 to C.B.E., Vannevar Bush Faculty Fellowship (N00014-20-1-2844), the Army Research Office through Grant W911NF-17-1-0462, AFOSR (FA9550-15-1-0334) and NSF through the University of Wisconsin MRSEC (DMR-1720415). Research at University of Nebraska-Lincoln was supported by the National Science Foundation through Materials Research Science and Engineering Center (NSF Grant No. DMR-1420645). The computational work at Pennsylvania State University by J. A. Z., V. G., and L. Q. C was partially supported by the US Department of Energy, Office of Science, Basic Energy Sciences, under Award Number DE-SC0020145 as part of the Computational Materials Sciences Program. J.A.Z. would also like to acknowledge support from 3M Inc. for support via a fellowship. Phase field and thermodynamic computations for this research were performed at the Pennsylvania State University's Institute for Computational and Data Sciences Advanced CyberInfrastructure (ICDS-ACI). Y. Y. and S. L. were partially supported by the US Department of Energy grant number DE-SC0012375 for the nonlinear optical characterization. The work at University of California, Irvine, was supported by the Department of Energy (DOE) under grant DESC0014430. TEM experiments were conducted using the advanced TEM facilities in the Irvine Materials Research Institute (IMRI) at the University of California, Irvine.

## Author Contributions

C.B.E., E.Y.T., L.Q.C., A.G., X.P., V.G. and T.T. supervised the experiments. K.E., J.W.L., S.R., H.L. and T.H.K. performed the sample growth and surface/structural characterizations. T.R.P. and E.Y.T. performed DFT calculations. B.W., J.A.Z. and L.Q.C. carried out phase field simulations. H.L. and A.G. performed PFM measurements. H.H., W.G. and X.P. carried out STEM measurements. S.L., K.E. and J.W.L. performed the device fabrication and ferroelectric measurements. Y.Y., S.L. and V.G. performed SHG measurements and analysis.

## Competition interests

The authors declare no competing financial interests.

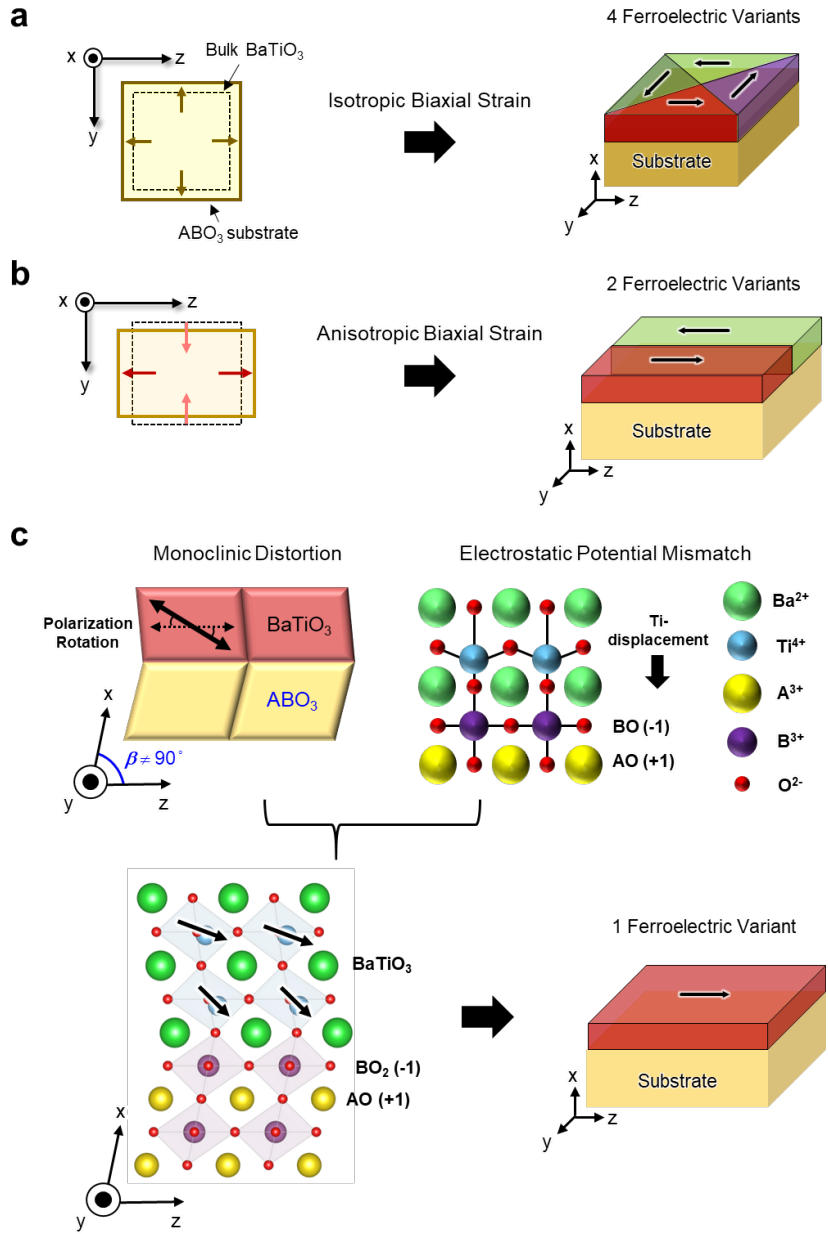

**Fig. 1 | Schematic diagram depicting the strategy employed to achieve single in-plane ferroelectric film. a**, BTO films grown under an isotropic biaxial tensile strain. **x**, **y**, and **z** correspond to the pseudocubic (pc) [100], [010], and [001] axes, respectively. Due to tensile strain, *c*-axis of BTO lies along the in-plane direction of the substrate. However, the isotropic geometry gives rise to form a multi-domain configuration where four ferroelectric variant exists. **b**, In the case of the BTO film grown under proper anisotropic biaxial strain, which is that the $b_{pc}$ and $c_{pc}$ of substrate are close to the *a/b*-axis and *c*-axis lattice parameters of tetragonal BTO, respectively, the BTO film is expected to have a polarization direction parallel to **z** directions, resulting in two ferroelectric variants. **c,** BTO films grown on the substrate which possesses proper in-plane lattice match, monoclinic nature, and polar planes (*e.g.*, $BO_2$-terminated $A^{3+}B^{3+}O_3$ substrates), one of two configurations can be stabilized. Note that the effect of electric field generated by polar planes is weaker and weaker as far away from the interface, which results in a single ferroelectric variant along in-plane direction.

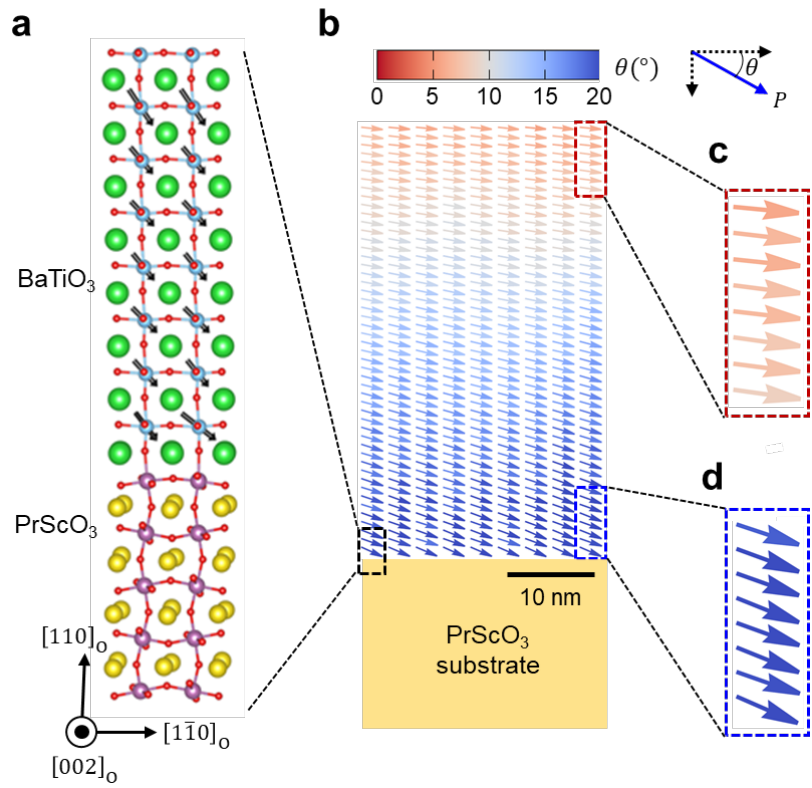

**Fig. 2 | Theoretical calculations of BTO/PSO heterostructure. a**, Calculated atomic structure of BTO/PSO $(110)_O$ heterostructure. **b**, The spatial polarization distribution of BTO film obtained from the phase field simulation. The film was assumed to be 50 nm thick with a fully coherent interface with respect to the substrate. Color bar reflects the polarization angle with respect to PSO $[1\bar{1}0]_O$. **c, d**, The out-of-plane polarization gradually decreases through the film thickness (from (**d**) to (**c**)), becomes negligible near the film surface, predicting single domain in-plane BTO film.

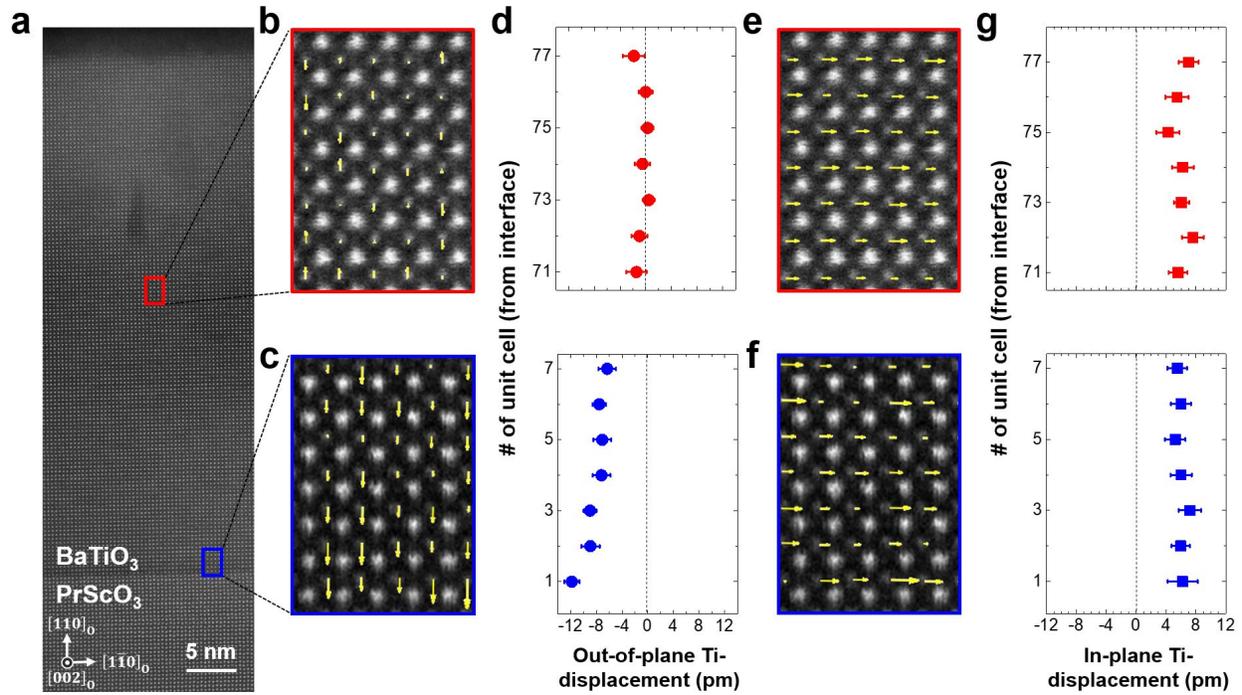

**Fig. 3 | STEM measurements of BTO/PSO heterostructure. a**, A low-magnification HAADF-STEM image of the BTO film with a zone axis of PSO $[002]_O$. **b**, **c**, High resolution images taken from selected areas (red and blue boxes in Fig. 3(**a**), respectively) shows out-of-plane component of Ti-displacement in the middle region (**b**), and near the interface region (**c**) of the BTO film, respectively. Note that the size of arrows corresponds to the amount of Ti-displacement. **d**, Average value of out-of-plane Ti-displacement in each layer of BTO from the interface. **e**, **f**, High resolution images showing in-plane component of Ti-displacement in the middle region (**e**), and near the interface region (**f**) of the BTO film. **g,** Average in-plane Ti-displacement in each layer of BTO from the interface. Both out-of-plane and in-plane Ti-displacements are extracted from the same regions in the BTO film. The error bars represent the 80% confidence intervals.

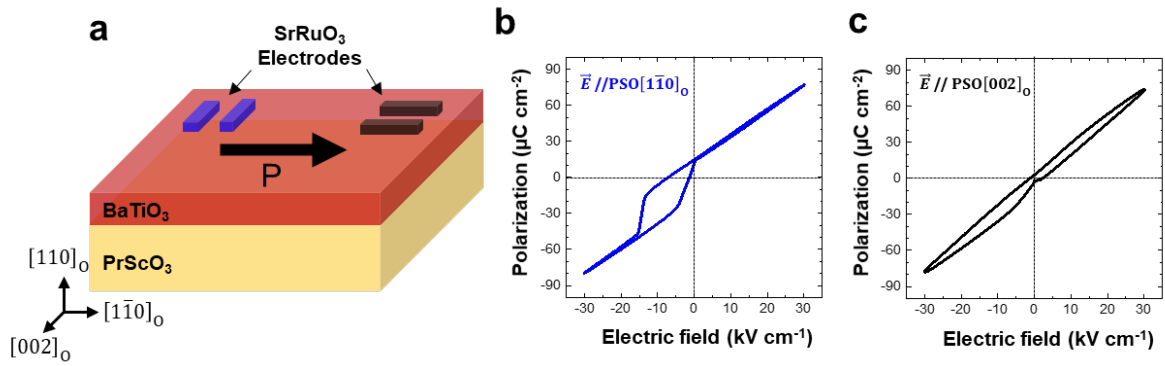

**Fig. 4 | In-plane polarization measurement versus electric field. a**, Schematics depicting parallel electrodes on the BTO/PSO sample. The 10-nm-thick SrRuO$_3$ patterns which are oriented along two different directions with 50-μm gap are used as electrodes for the measurements. **b, c**, Room temperature hysteresis loop measurements for two parallel electrodes arrangements where the electric field is applied parallel to PSO $[1\bar{1}0]_O$ (**b**), and PSO $[002]_O$ (**c**) directions, respectively.

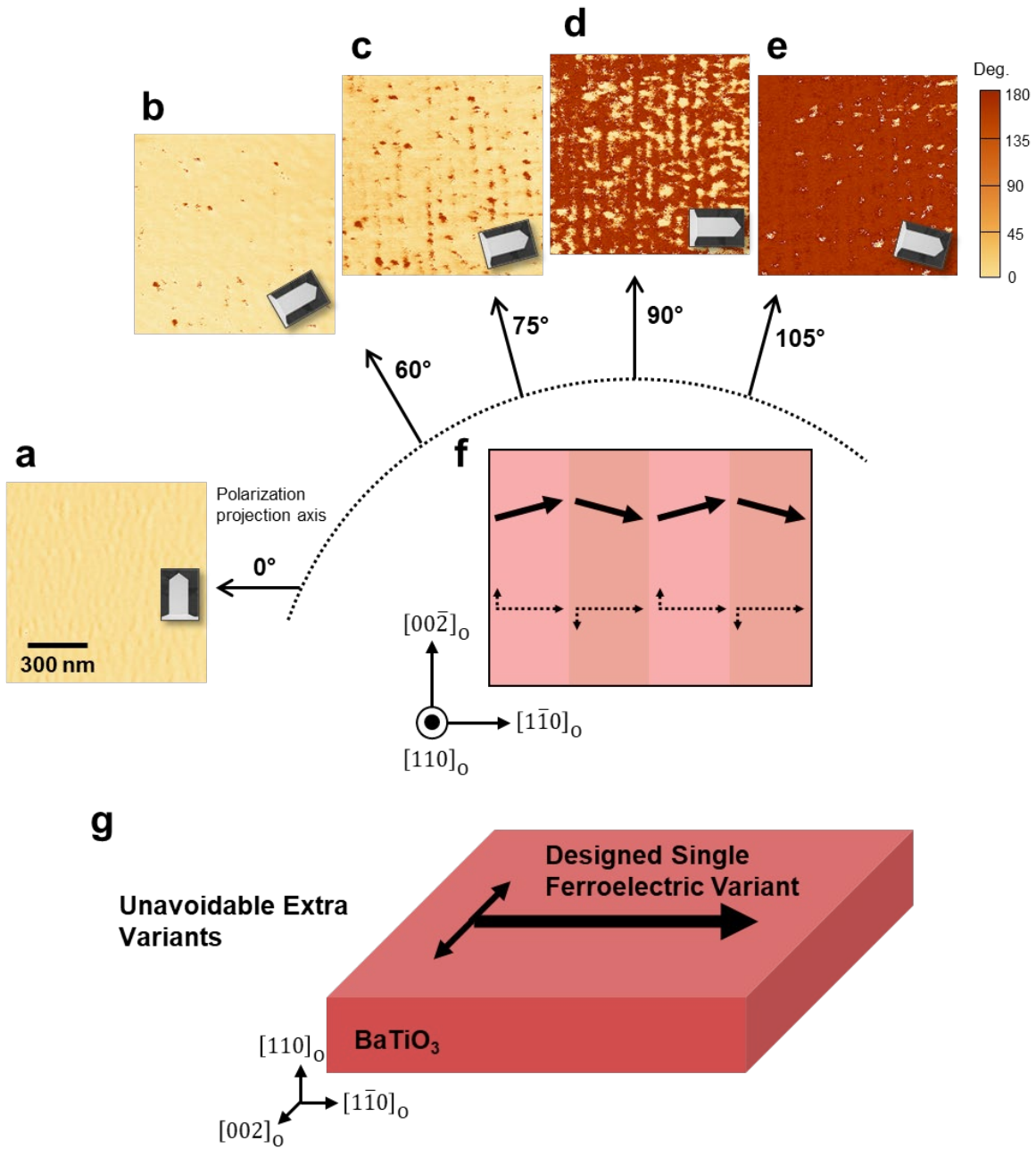

**Fig. 5 | Angular dependent lateral PFM images and an overall in-plane polarization configuration. a-e**, Lateral PFM phase images with various relative sample-cantilever arm angles of 0° (**a**), 60° (**b**), 75° (**c**), 90° (**d**), and 105° (**e**). The inset shows the orientation of the cantilever with respect to the crystallographic directions of the PSO substrate. Note that the angle is defined by the angle between the polarization projection axis and PSO $[1\bar{1}0]_O$ direction. **f**, In-plane domain structure constructed from PFM images of **a-e**. The dashed arrows show two orthogonal in-plane components of in-plane polarization (solid arrows), **g,** Overall configuration of in-plane polarization in the BTO film. Designed single ferroelectric variant is achieved along PSO $[1\bar{1}0]_O$ direction, while there are unavoidable extra variants along PSO $\langle 002 \rangle_O$ directions due to absence of suitable single crystal substrates.

Supplementary Information for

# In-plane quasi-single-domain BaTiO$_3$ *via* interfacial symmetry engineering

J. W. Lee[1,9], K. Eom[1,9], T. R. Paudel[2,3], B. Wang[4], H. Lu[2], H. Huyan[5], S. Lindemann[1], S. Ryu[1], H. Lee[1], T. H. Kim[1], Y. Yuan[4], J. A. Zorn[4], S. Lei[4], W. Gao[5], T. Tybell[6], V. Gopalan[4], X. Pan[5,7,8], A. Gruverman[2], L. Q. Chen[4], E. Y. Tsymbal[2], and C. B. Eom[1*]

[1]Department of Materials Science and Engineering, University of Wisconsin-Madison, Madison, Wisconsin 53706, USA; [2]Department of Physics and Astronomy & Nebraska Center for Materials and Nanoscience, University of Nebraska, Lincoln, Nebraska 68588, USA; [3]Department of Physics, South Dakota School of Mines and Technology, Rapid City, South Dakota 57701, USA; [4]Department of Materials Science and Engineering, The Pennsylvania State University, University Park, Pennsylvania 16802, USA; [5]Department of Materials Science and Engineering, University of California, Irvine, California 92697, USA; [6]Department of Electronic Systems, Norwegian University of Science and Technology, 7491 Trondheim, Norway; [7]Department of Physics and Astronomy, University of California, Irvine, California 92697, USA; [8]Irvine Materials Research Institute, University of California, Irvine, California 92697, USA; [9]These authors contributed equally to this work.

* Corresponding author. Email: eom@engr.wisc.edu

**Supplementary Figures**

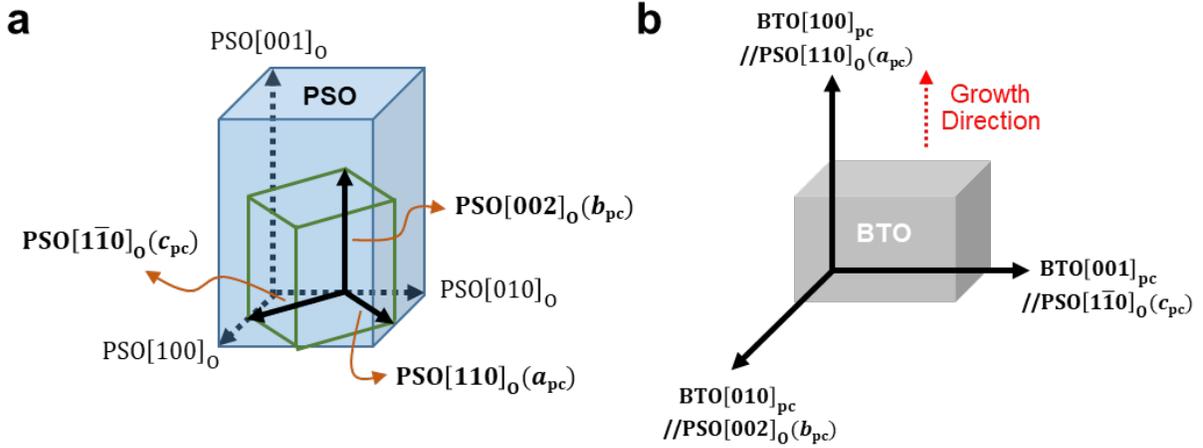

**Supplementary Figure 1. Axis notation for BTO and PSO. a**, Schematic diagram showing PSO crystal structure in orthorhombic unit cell. The inner cube is the pseudocubic unit cell. **b**, The epitaxial arrangement of $(100)_{pc}$-oriented BTO on PSO $(110)_O$ substrate. The growth direction is oriented along the PSO $[110]_O$ (or BTO $[100]_{pc}$). Note that the BTO $[001]_{pc}$ and BTO $[100]_{pc}$ directions are not exactly parallel with the $c_{pc}$ (PSO $[1\bar{1}0]_O$) and $a_{pc}$ (PSO $[110]_O$) directions, respectively, due to the monoclinic nature of the pseudocubic PSO unit cell.

**Supplementary Note 1: DFT calculations**

    DFT calculations were performed with two different initial polarizations of BTO with corresponding two monoclinic tilt directions: the opposite tilt direction (Fig. 2a) and the same tilt direction (Supplementary Figure 2a). The results show that the BTO/PSO with the opposite tilting direction (Fig. 2a) has lower energy than that of the same tilting direction (Supplementary Figure 2a) by 18.6 mJ/m$^2$. The origin of such a lower energy state is associated to the interfacial Sc-O-Ti network. It should be noted that the bulk PSO has Sc-O-Sc bonding angle of ~147° with ScO$_6$ octahedral rotation pattern ($a^-a^-c^+$ in Glazer's notation)[1-3], while bulk BTO does not have any rotation/tilt at room temperature[4]. In the BTO/PSO heterostructure, octahedral tilts can propagate into the BTO layers near the interface region[4]. The calculated Sc-O-Ti angles are 171°, 173° in the case of the opposite tilt direction (Supplementary Figure 2b) and 176°, 168° in the case of the same tilt direction (Supplementary Figure 2c), respectively. Relatively large angle difference (~8°) between two adjacent Sc-O-Ti bond angles in the same tilt direction is likely to result in the higher energy state.

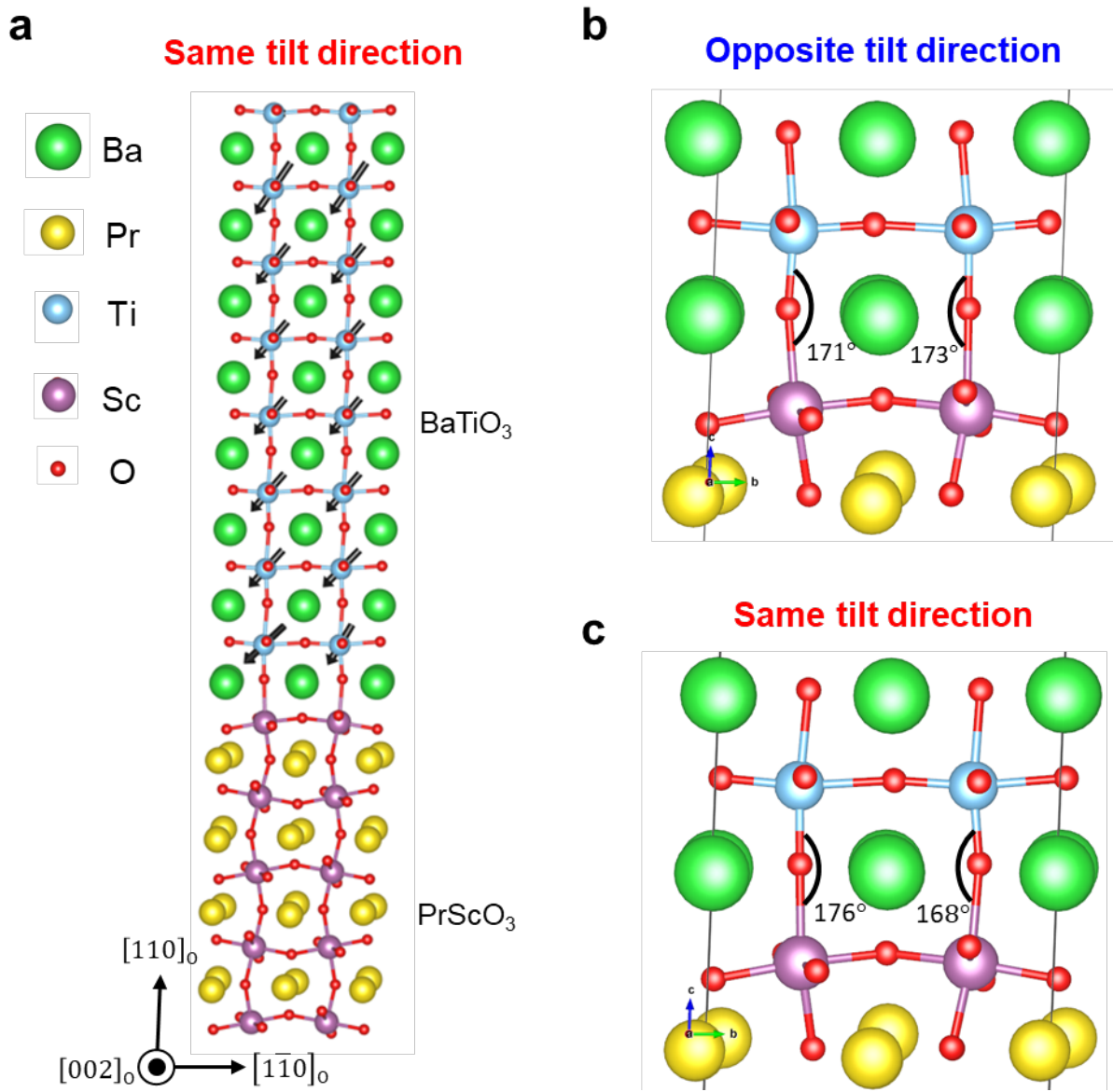

**Supplementary Figure 2. Density functional theory calculations for BTO/PSO heterostructure. a**, The relaxed atomic structure by DFT calculation in the case of the same tilt direction. The initial Ti-displacement was set to the diagonal direction (i.e., combination of PSO $[\bar{1}\bar{1}0]_O$ and $[\bar{1}10]_O$). **b**, **c**, The interfacial atomic structure with the bond angle between Sc-O-Ti in the case of (**b**) the opposite tilt direction and (**c**) the same tilt direction, respectively.

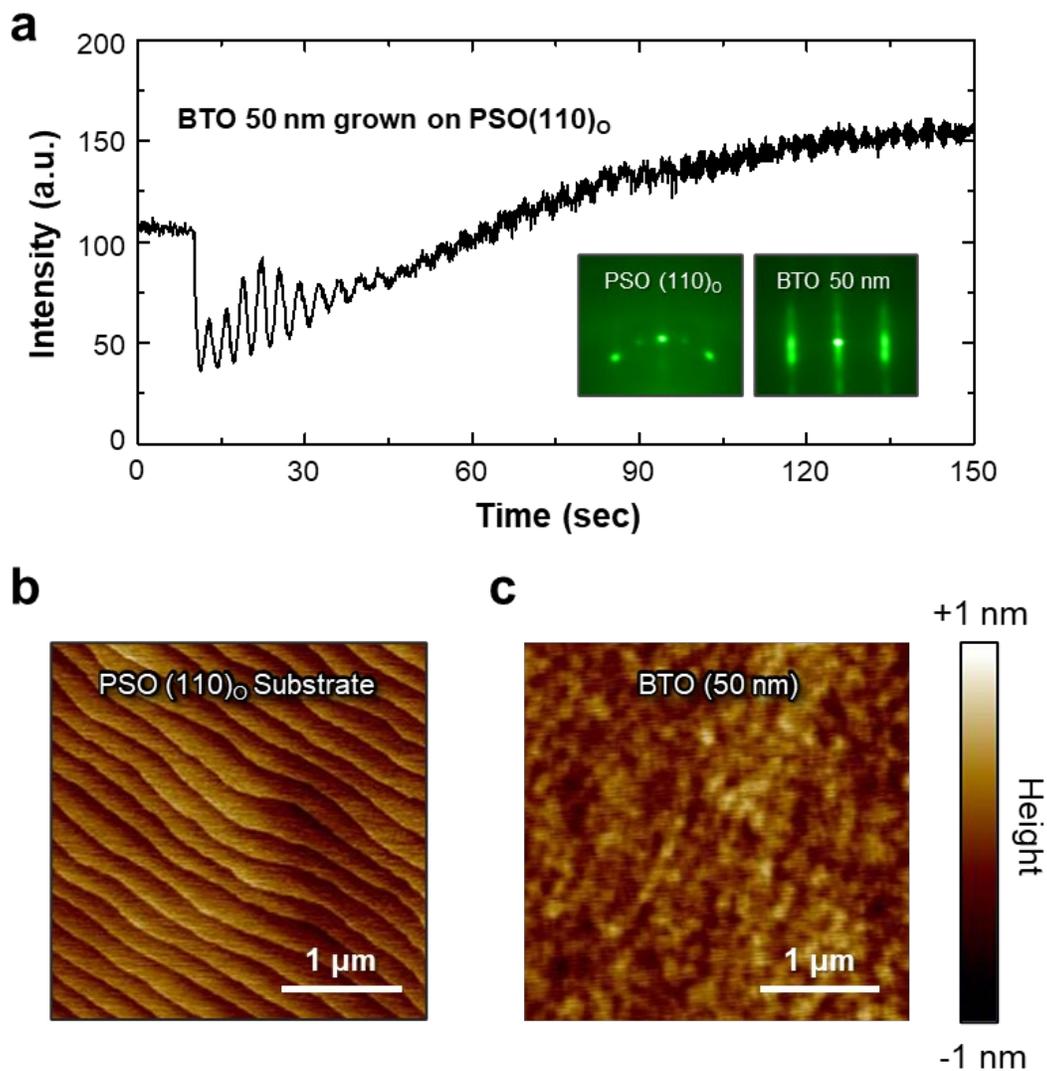

**Supplementary Figure 3. In-situ RHEED observation and surface topography. a**, RHEED oscillations for the growth of BTO. The data from initial growth (~140 seconds) is represented. The insets show the RHEED patterns of a PSO (110)$_O$ substrate and a 50-nm-thick BTO film. **b**, **c**, AFM topography images of a treated PSO (110)$_O$ substrate (**b**) and 50-nm-thick BTO film on PSO (110)$_O$ substrate (**c**).

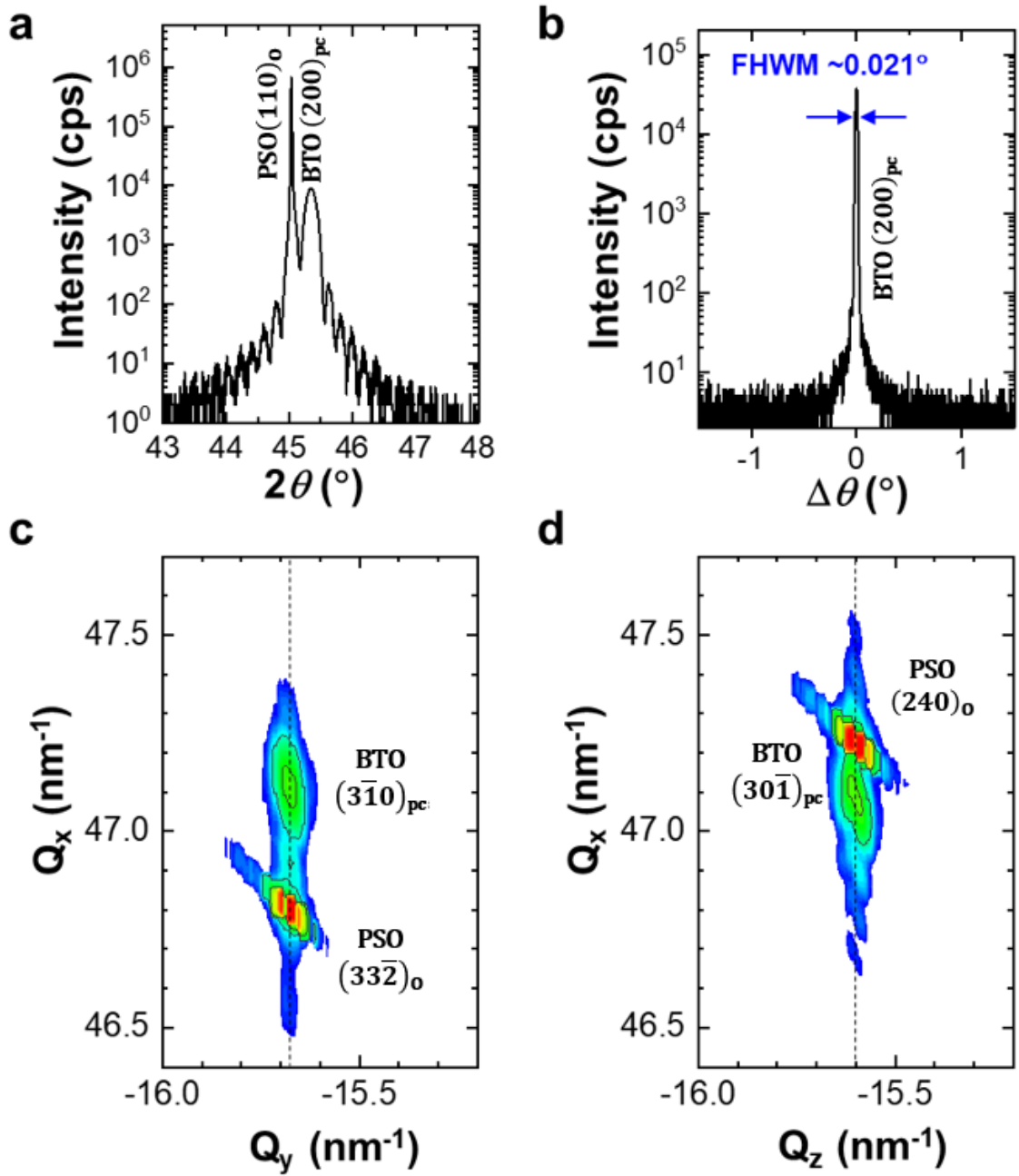

**Supplementary Figure 4. X-ray diffraction patterns of epitaxial BTO films grown on PSO $(110)_O$ substrates. a, b,** (**a**) Out-of-plane $\theta$-$2\theta$ measurement, (**b**) a BTO $(200)_{pc}$ rocking curve of the BTO film. **c, d,** Reciprocal space maps of the BTO film around PSO $(33\bar{2})_O$ (**c**) and PSO $(240)_O$ (**d**) Bragg peaks, respectively.

**Supplementary Note 2: The overall polarization state of the BTO film**

In addition to interface and middle region of the BTO film, the polarization state in the top region of the BTO film (close to the surface) is also shown in the Supplementary Figure 5. It is revealed that in-plane polarization direction of BTO film is toward PSO $[1\text{-}10]_O$ (Supplementary Figure 2c, which is in agreement with the middle region (Fig. 3e) and near interface region (Fig. 3f). The average Ti displacement is estimated to 6.9 ± 1.3 pm, which is comparable to the values in the middle region of the film (6.0 ± 1.1 pm) and BTO/PSO interface region (6.0 ± 1.0 pm) within the error range. This uniform in-plane Ti displacement is mainly attributed to the coherent nature of BTO film on PSO substrate (Supplementary Figure 4c, d). In addition, there is a tiny amount of out-of-plane component pointing upward (toward top surface; 1.9 ± 1.1 nm). We think that this tiny amount of out-of-plane component pointing upward is presumably originated from external factors, e.g, the electronic boundary condition formed by surface adsorbates such as $H_2O$[5].

The overall displacement maps near interface, middle and top regions of the BTO film are also shown in Supplementary Figure 6. It is clearly seen that the in-plane polarization direction is along PSO $[1\text{-}10]_O$ in all images. While there is a dominant downward polarization (out-of-plane polarization toward PSO $[\text{-}1\text{-}10]_O$ direction) near interface due to interfacial electrostatic potential, the randomly-oriented out-of-plane polarization is observed in the middle region.

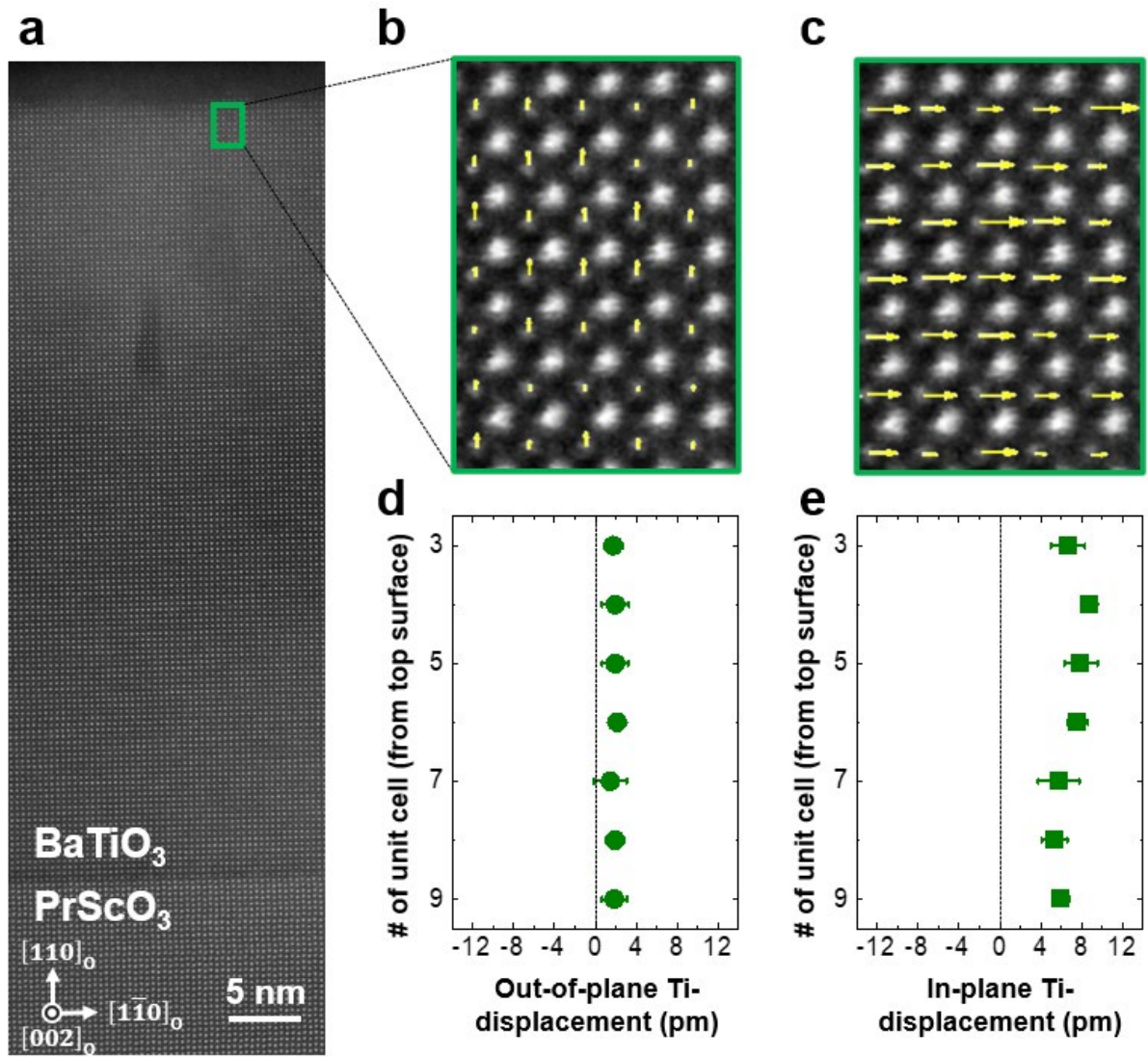

**Supplementary Figure 5. STEM measurements of the top region of BTO film on PSO (110)$_O$ substrate.** **a**, A low magnification HAADF-STEM image of BTO film with a zone axis of PSO [002]$_O$. **b, c**, High resolution images showing (**b**) out-of-plane and (**c**) in-plane components of Ti displacement in the top region of the BTO film, which is marked in green color in (**a**). Note that the size of arrows corresponds to the amount of Ti displacement. **d, e**, Ti displacement of (d) in-plane and (e) out-of-plane component as a function of BTO unit cell from the top surface.

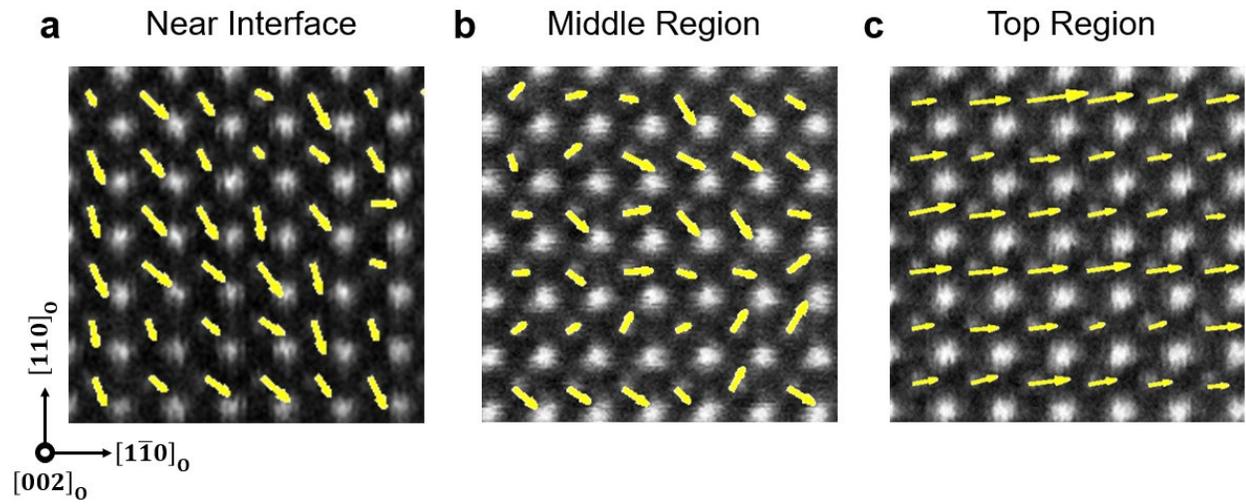

**Supplementary Figure 6. Overall polarization state of the BTO film on PSO (110)$_O$ substrate. a-c**, A high resolution HAADF-STEM images with a zone axis of PSO [002]$_O$ showing overall polarization low magnification HAADF-STEM image of BTO film at (**a**) the interface region, (**b**) the middle of the film, and (**c**) the top region of the film, respectively.

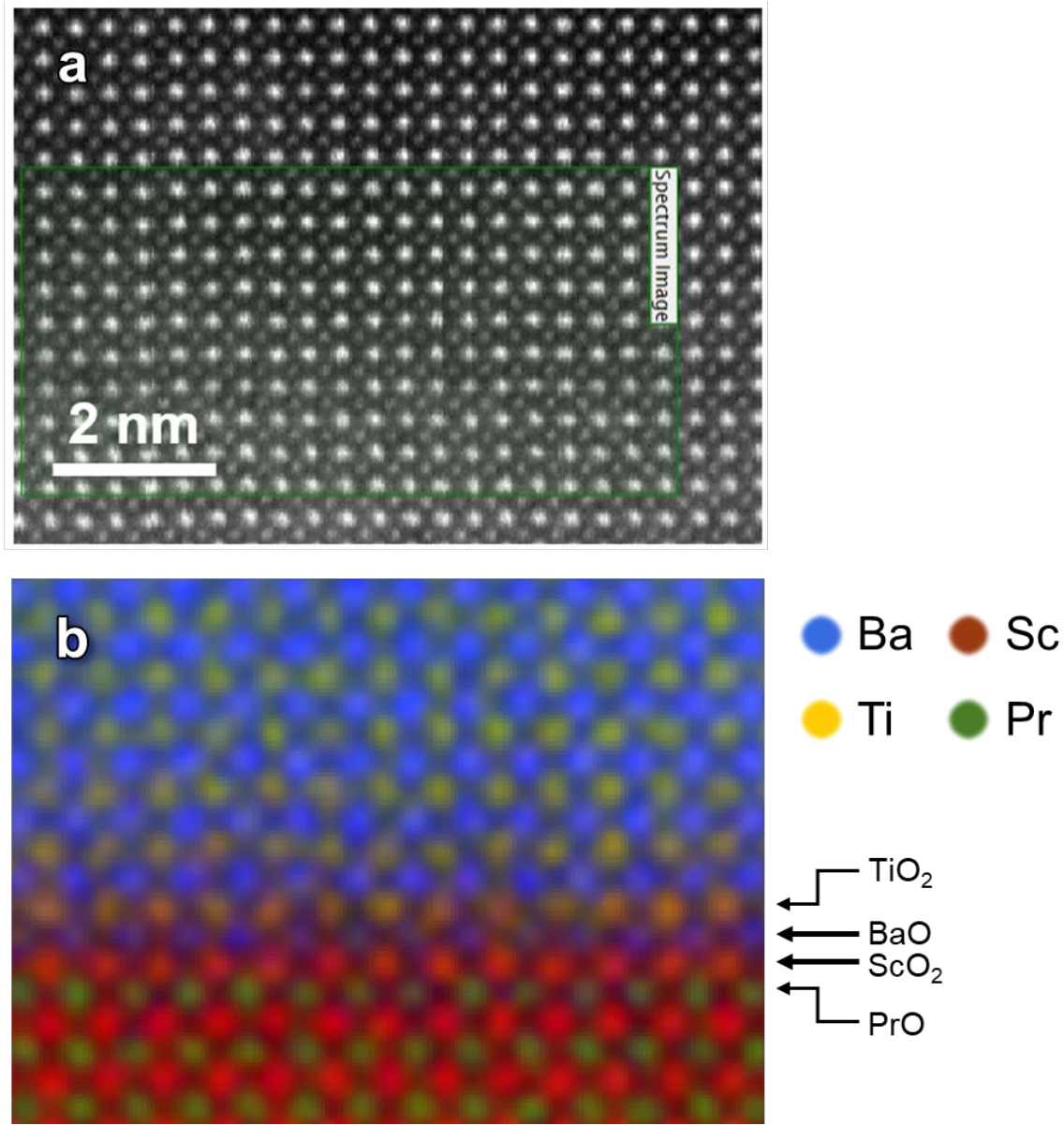

**Supplementary Figure 7. Compositional analysis near the interface region between BTO and PSO**. **a**, HAADF-STEM images with a zone axis of PSO $[002]_O$. **b**, Energy dispersive spectroscopy (EDS) elemental mapping images indicating $ScO_2$-termination of PSO $(110)_O$ substrate.

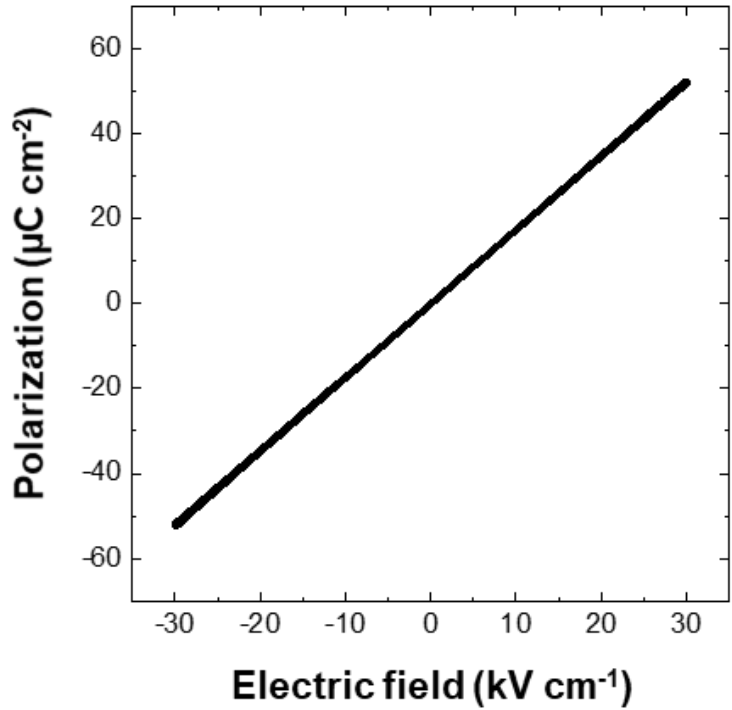

**Supplementary Figure 8. In-plane polarization measurement versus electric field.** Room temperature hysteresis loop for two parallel electrodes on the bare PSO substrate. The electric field is applied parallel to PSO $[002]_O$. There is a linear relationship between polarization and applied electric field, indicating that the overall positive slope in the BTO/PSO sample shown in Fig. 4b and c is originated from the PSO substrate.

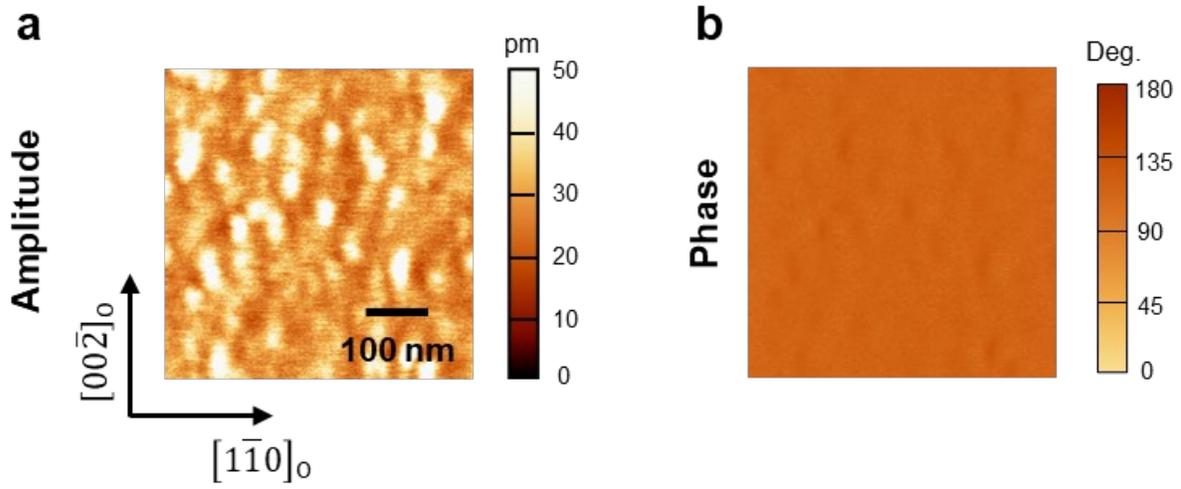

**Supplementary Figure 9. Vertical PFM images of the BTO film on PSO substrate**. **a**, amplitude image. **b**, phase image.

## Supplementary Note 3: Domain structure along the PSO $[002]_O$ and its origin

Angular dependent lateral PFM imaging of the domain structures are shown in Supplementary Figure 10. In the lateral PFM mode, signal is sensitive only to the polarization perpendicular to the cantilever arm, so that the measured domain structure represents a projected polarization along the axis perpendicular to the cantilever arm (polarization projection axis). A physical rotation of the sample relative to the cantilever arm was adopted in Supplementary Figure 10 while imaging in the exact same area, and the polarization directions sensitive to the specific sample-cantilever arm angle are represented by blue arrows (Supplementary Fig. 10f-j). First, the PFM image taken with polarization projection axis along PSO $[1\bar{1}0]_O$ (referred to as 0° angle) exhibit no contrast, indicating a single domain state (Supplementary Fig. 10f). By rotating the relative sample-cantilever arm angle, the small portion of contrast begins to appear as the angle reaches 75° (Supplementary Fig. 10h). The PFM image taken at 90° (with polarization projection along the PSO $[002]_O$ direction) shows both up and down domain (Supplementary Fig. 10i). The data with 105° represent similar tiny domain feature with to that of 75° (Supplementary Fig. 10j). From this angular dependent PFM imaging results, we conclude that the in-plane polarization direction is mainly toward PSO $[1\bar{1}0]_O$, with small polarization tilting angles toward PSO $[002]_O$ and PSO $[00\bar{2}]_O$ of a magnitude less than 15° in majority of the area. Supplementary Figure 11 shows the reconstructed image of the actual polarization orientation (with the angle relative to PSO $[1\bar{1}0]_O$ axis) from analysis of PFM images, and a histogram distribution of the polarization angles within the inspected area.

This non-zero polarization along PSO $[002]_O$ direction is also supported by high magnification HAADF-STEM Images of a BTO film along the PSO $[1\bar{1}0]_O$ zone-axis (Supplementary Figure 12). Ti displacements toward both PSO $[002]_O$ and $[00\bar{2}]_O$ direction are observed (Supplementary Figure 12c, d). It should be noted that the magnitude of Ti displacement along the PSO $[002]_O$ and $[00\bar{2}]_O$ is much smaller than that of along the PSO $[1\bar{1}0]_O$ (Fig. 3e, f). In addition, the overall in-plane domain structure is consistent with phase field simulation (Supplementary Figure 13b). When BTO is coherently grown on PSO $(110)_O$, c-axis (long axis) of bulk tetragonal BTO is along PSO $\langle 1\bar{1}0 \rangle_O$ under compressive strained (Supplementary Figure 13a) while a/b-axis (short axis) of bulk tetragonal BTO is along PSO $\langle 002 \rangle_O$ under tensile strained state. This may lead to form zigzag patterns of in-plane domain structure as shown in Supplementary Figure 13b and 14. This explanation is further supported by simulation results performed under absence of tensile strain along the PSO $\langle 002 \rangle_O$ direction (Supplementary Figure 13c). The result shows a pure single domain structure where polarization direction is perfectly toward to PSO $[1\bar{1}0]_O$ direction (Supplementary Figure 13d and 15), indicating the validity of our strategy.

It should be noted that the domain pattern tends to become more complex with additional extrinsic factors at play. Specifically, when PFM analysis is performed closer to the edge of the sample, a polydomain structure arises due to the large-scale periodic changes of the polarization direction along PSO $[\bar{1}10]_O$ direction (Supplementary Figure 16). It is feasible to assume that the elastic strain conditions are quite different at the vicinity (in the range of several hundreds of micrometers) of the substrate edge in comparison to the substrate interior area. In this case, more complex patterns of strain anisotropy and lattice distortion would create conditions for domain periodicity along different crystallographic directions. Detail analysis of the effect of such extrinsic factors on the in-plane polarization alignment are underway.

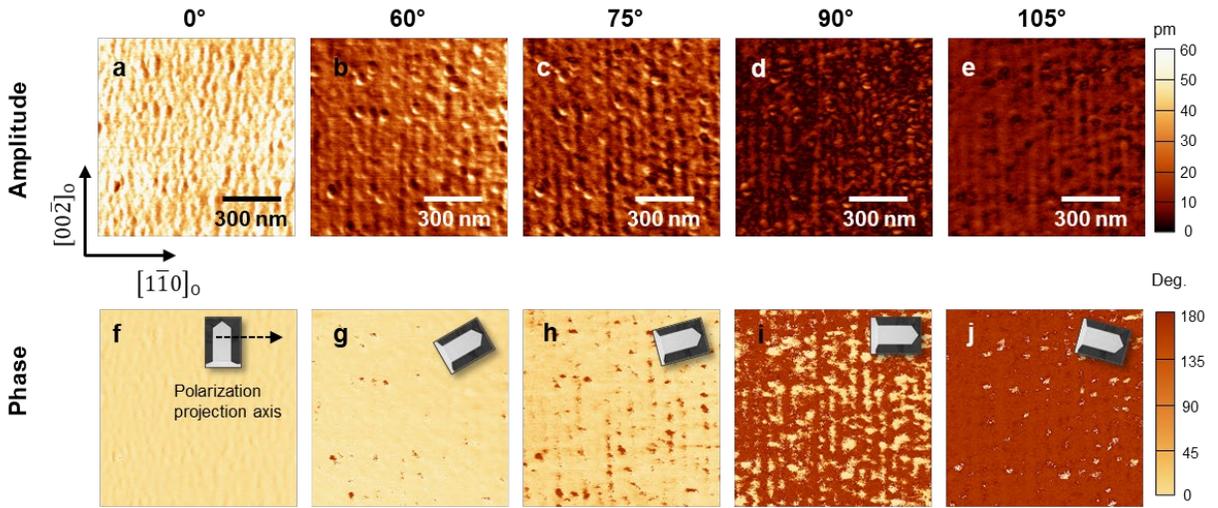

**Supplementary Figure 10. Angular dependent LPFM (a-e) amplitude and (f-j) phase images of the BTO film on PSO substrate at the same location**. The relative sample-polarization projection axis angles: 0° (used as a reference) (**a, f**), 60° (**b, g**), 75° (**c, h**), 90° (**d, i**), 105° (**e, j**).

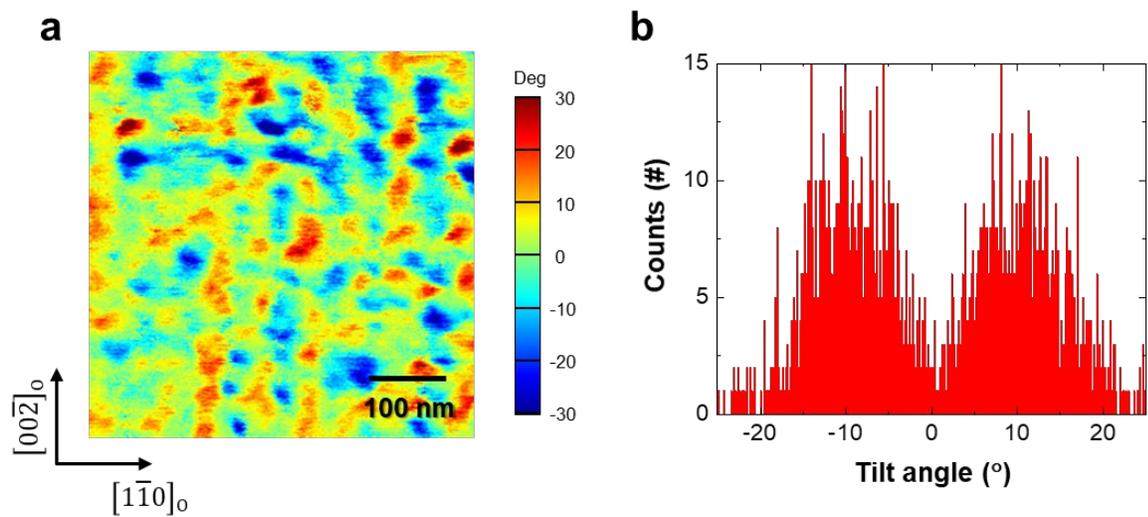

**Supplementary Figure 11. Actual polarization angles relative to PSO $[1\bar{1}0]_O$ direction**. **A**, Reconstructed polarization orientation map relative to PSO $[1\bar{1}0]_O$ direction from the PFM images. **B**, Histogram distribution of the polarization tilting angle relative to PSO $[1\bar{1}0]_O$ direction.

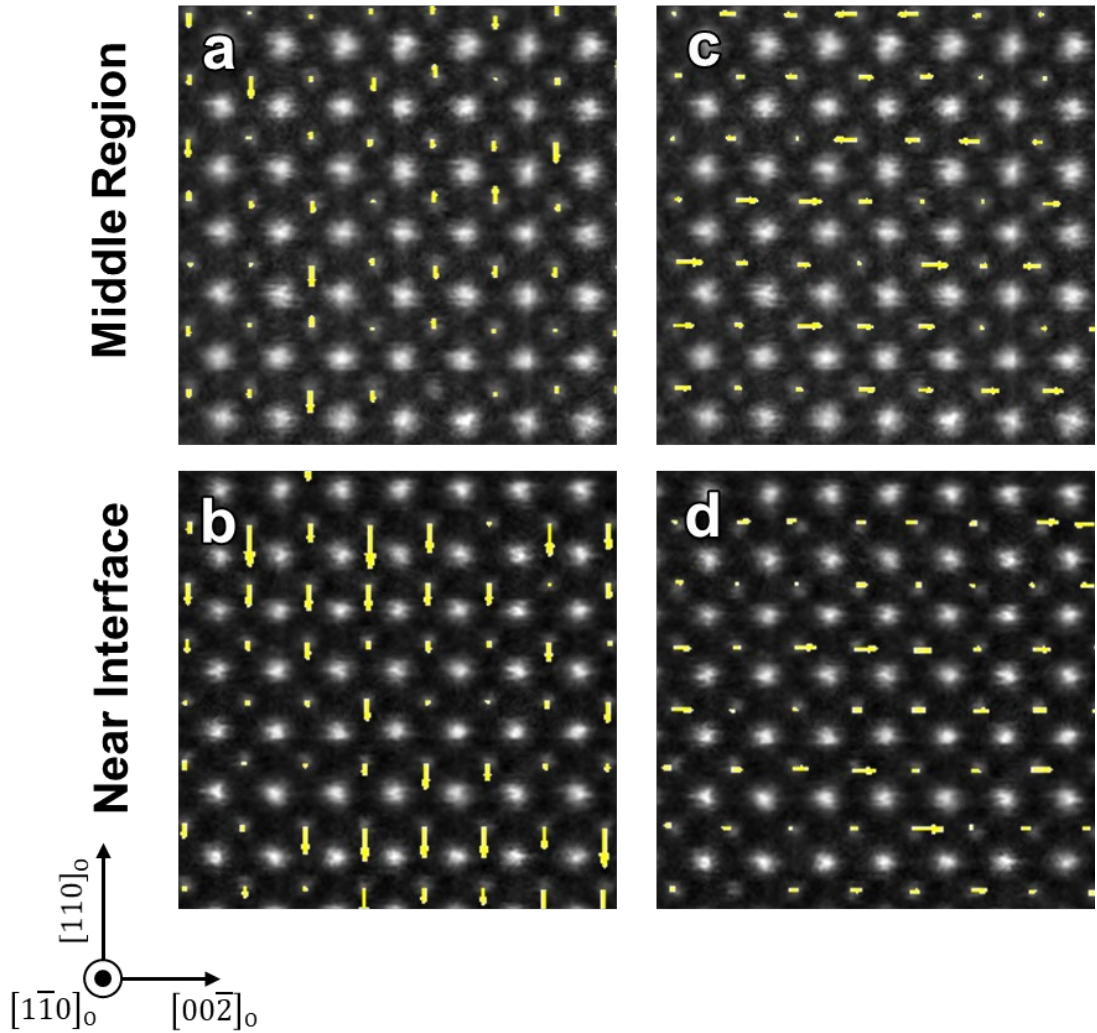

**Supplementary Figure 12. HAADF-STEM images of BTO film with a zone axis of PSO $[1\bar{1}0]_o$. a, b,** Out-of-plane component of Ti-displacement in the middle region (**a**) and near interface region (**b**) of the BTO film, respectively. **c, d,** In-plane component of Ti-displacement in the middle region (**c**) and near interface region (**d**) of the BTO film, respectively.

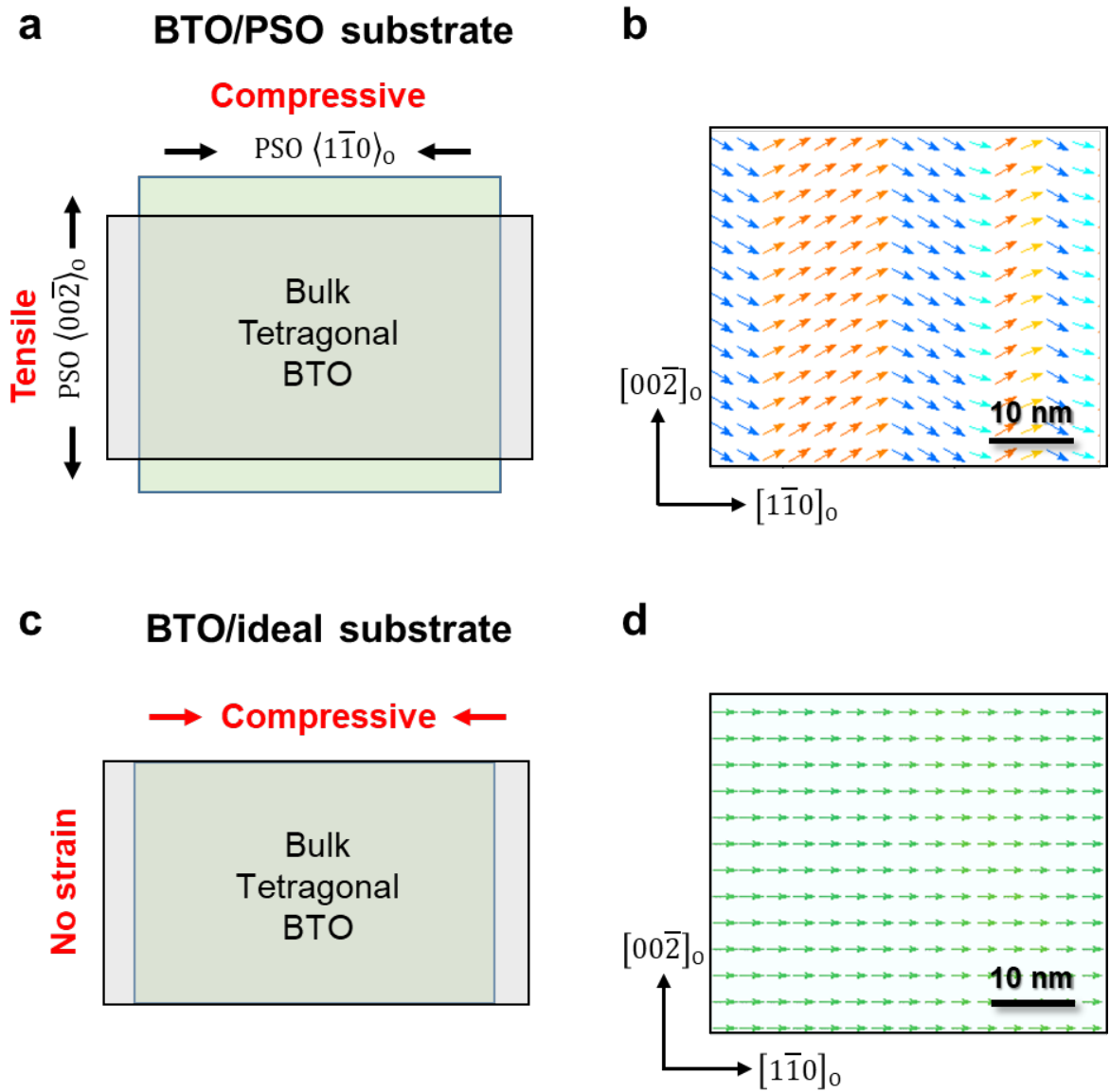

**Supplementary Figure 13. In-plane domain structure simulated by phase field simulation**. **a**, The schematic of the strain relationship between bulk tetragonal BTO and PSO $(110)_O$ substrate. **b,** Calculated zigzag domain patterns in BTO/PSO $(110)_O$ substrate. **c**, The schematic of BTO grown on the ideal substrate. **d**, Calculated single in-plane ferroelectric domain in BTO/ideal substrate.

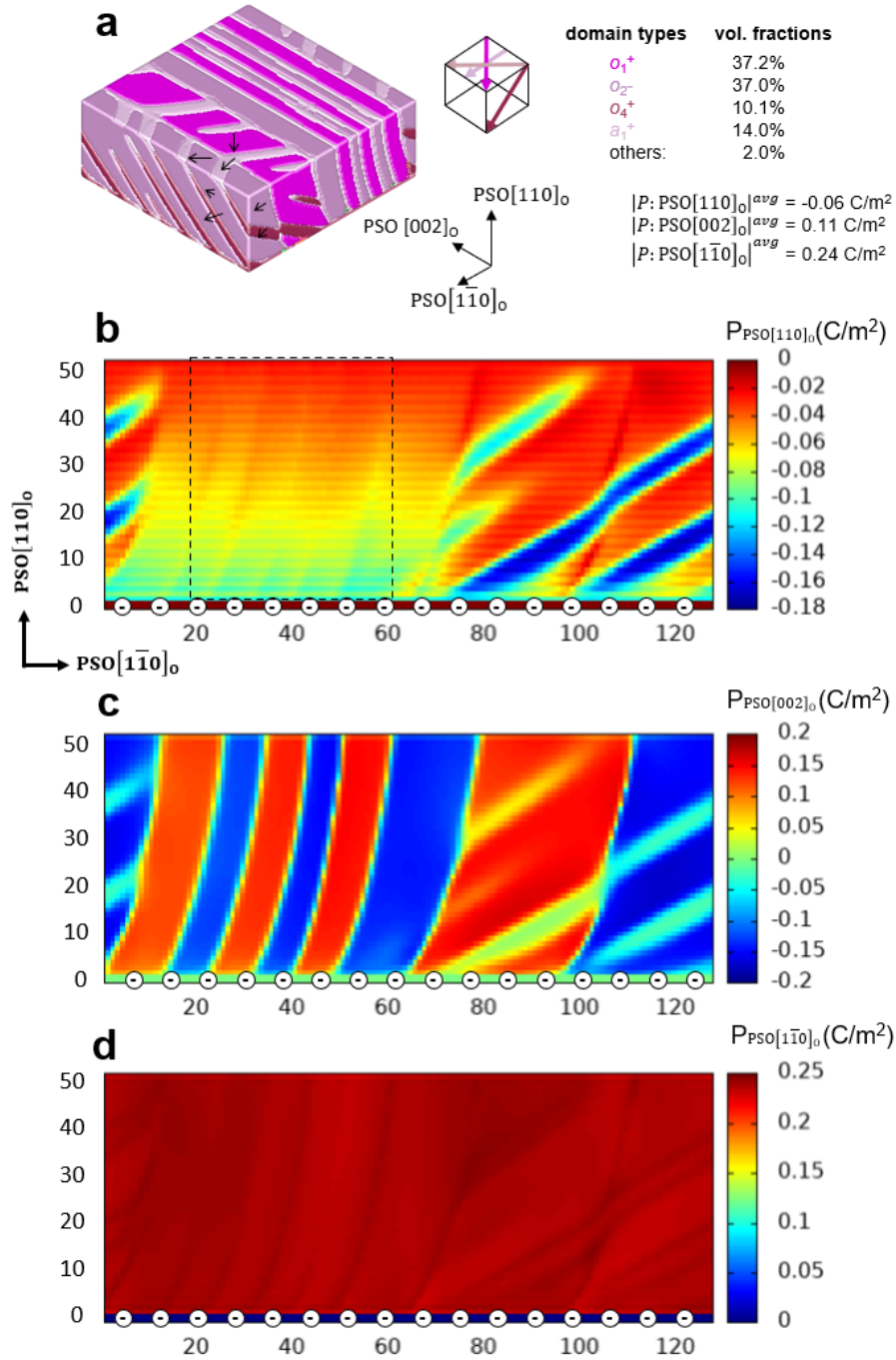

**Supplementary Figure 14. Phase-field simulation of the 50 nm BTO thin film on PSO $(110)_O$ substrate. a,** The three-dimensional domain structure of the BTO thin film at equilibrium. The colors represent different domain variants as determined by the direction of polarization vectors, as shown in the legend. The volume fraction of each domain variant and the averaged polarization magnitude along the three orthogonal direction of the system are given. **b-d,** The mappings of polarization vectors for **P**:PSO $[110]_O$ (**b**), **P**:PSO $[002]_O$ (**c**), and **P**:PSO $[1\bar{1}0]_O$ (**d**) components across a two-dimensional section in the PSO $[1\bar{1}0]_O - [110]_O$ plane. The negatively charged interfacial layer is schematically drawn in (**b-d**). The dashed rectangular region in (**b**) denotes the selected region shown in Figure 2 of the main text.

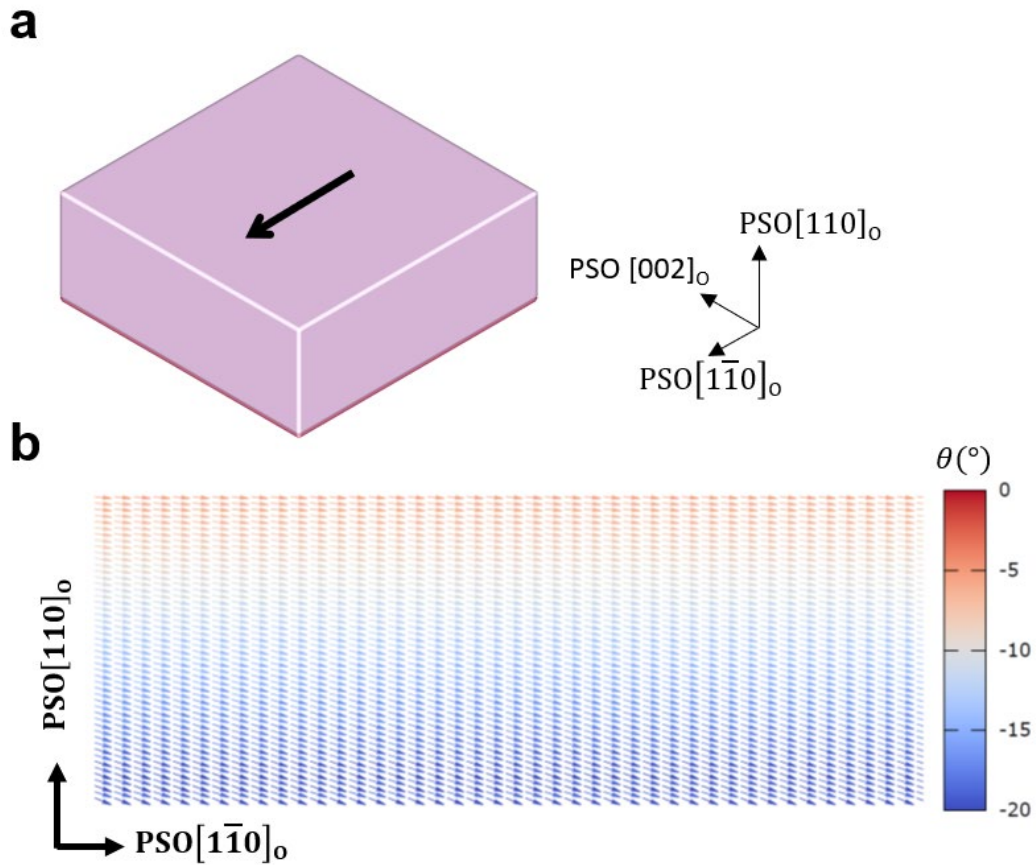

**Supplementary Figure 15. Phase-field simulation of the 50 nm BTO thin film on ideal substrate.** (a) The 3D domain structure of the BTO thin film at equilibrium, which shows a uniform $a_1^+$ domain with $P \parallel$ PSO $[110]_O$ (b) The surface view and (c) the section view of the distribution of polarization vectors colored by the rotation angle of polarization with respect to the PSO $[110]_O$ direction.

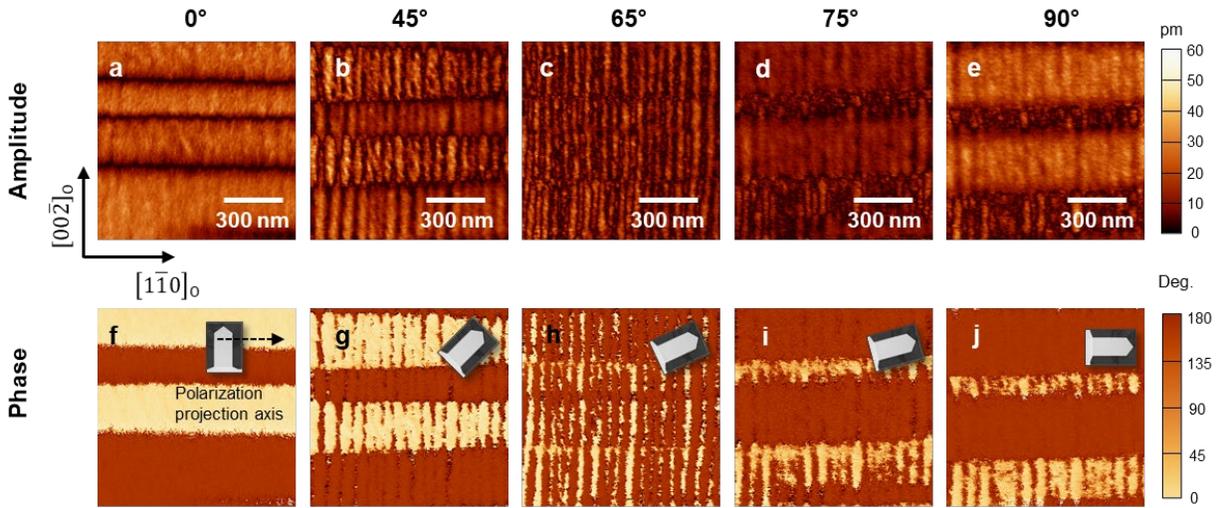

**Supplementary Figure 16. Angular dependent LPFM (a-e) amplitude and (f-j) phase images of the BTO film on PSO substrate near the edge region**. The relative sample-polarization projection axis angles: 0° (used as a reference) (**a, f**), 45° (**b, g**), 65° (**c, h**), 75° (**d, i**), 90° (**e, j**).

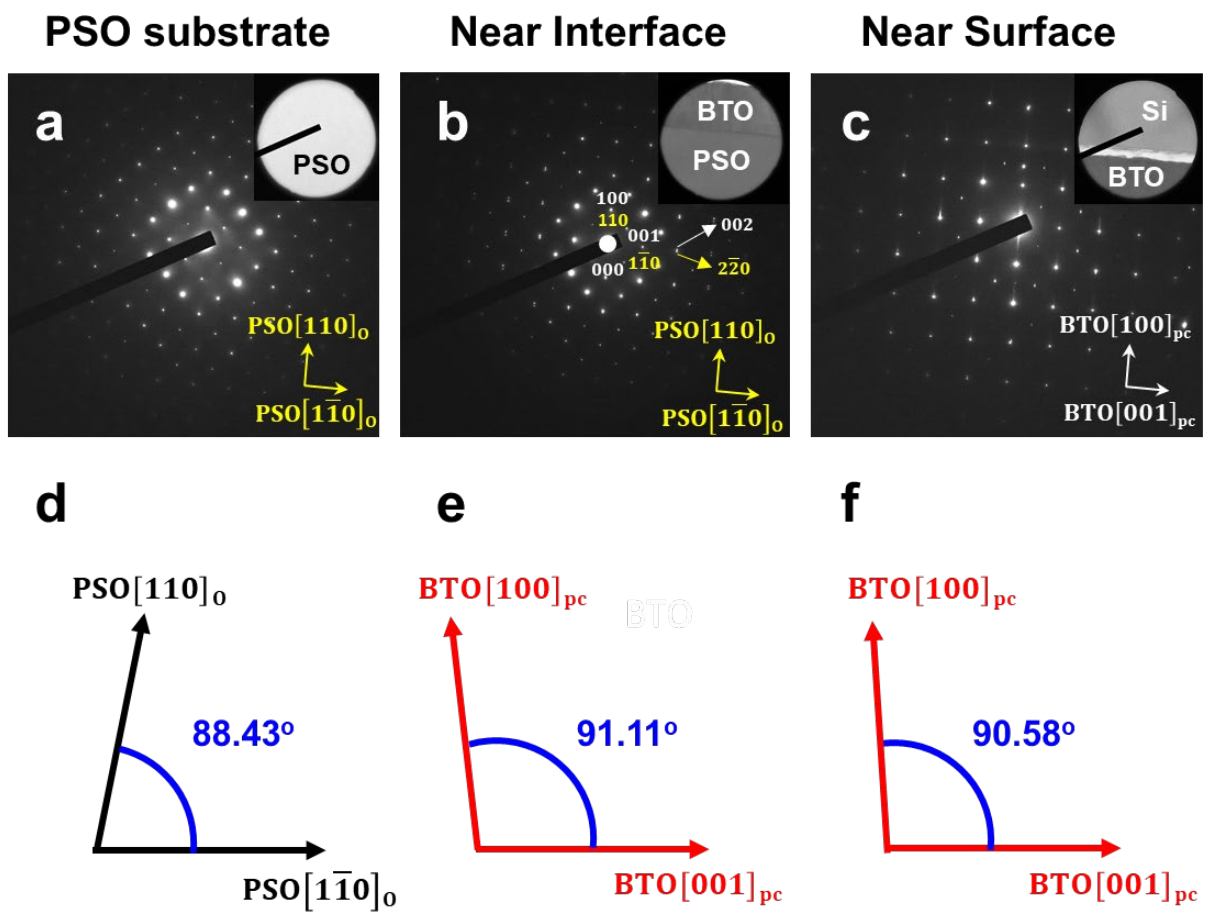

**Supplementary Figure 17. The diffraction patterns and tilt angles of PSO and BTO with a zone axis of PSO $[002]_O$**. **a-c**, Diffraction patterns for PSO substrate (**a**), at the interface between BTO/PSO (**b**), and near the top surface of BTO film (**c**). **d–f,** The angle between in-plane and out-of-plane axis of PSO substrate (**d**), BTO near the interface region (**e**), BTO near the surface region (**f**). Note that monoclinic tilting is decreased in BTO near surface region, indicating the structural relaxation.

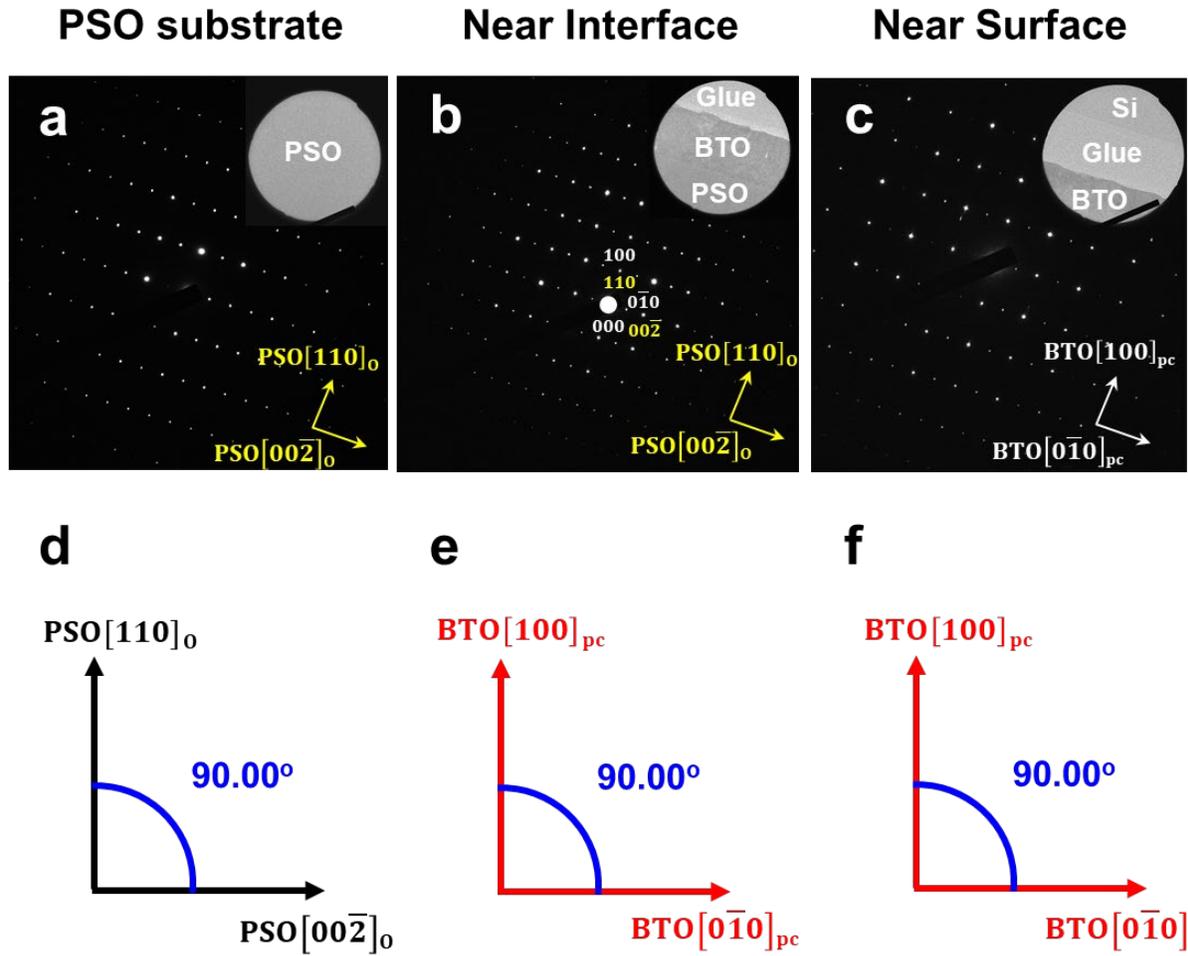

**Supplementary Figure 18. The diffraction patterns and tilt angles of PSO and BTO with a zone axis of PSO $[1\bar{1}0]_O$. a-c**, Diffraction patterns for PSO substrate (**a**), at the interface between BTO/PSO (**b**) and near the top surface of BTO film (**c**). **d–f**, The angle between in-plane and out-of-plane axis of PSO substrate (**d**), BTO near the interface region (**e**), BTO near the surface region (**f**). Note that there is no monoclinic tilt in BTO both near interface and top surface regions.

## Supplementary Note 4: The strain analysis of the BTO film on PSO (110)$_O$ substrate

    Based on our model, there is a strain gradient in our BTO film on PSO (110)$_O$ substrate with respect to thickness direction. This strain gradient is also observable in geometric phase analysis (GPA) from TEM data. As shown in Supplementary Figure 19(b), it is clear that BTO film is under higher tensile strain state near interface, which is relaxed within 10 nm (Supplementary Figure 19(d)). This is consistent with our model where electric field across the interface can affect the interface region of BTO causing downward polarization, and then might be weaker in the middle region of the films. On the other hand, there is no strain gradient along in-plane direction (Supplementary Figure 19(c) and (e)), because our BTO film is fully coherent with PSO (110)$_O$ substrate.

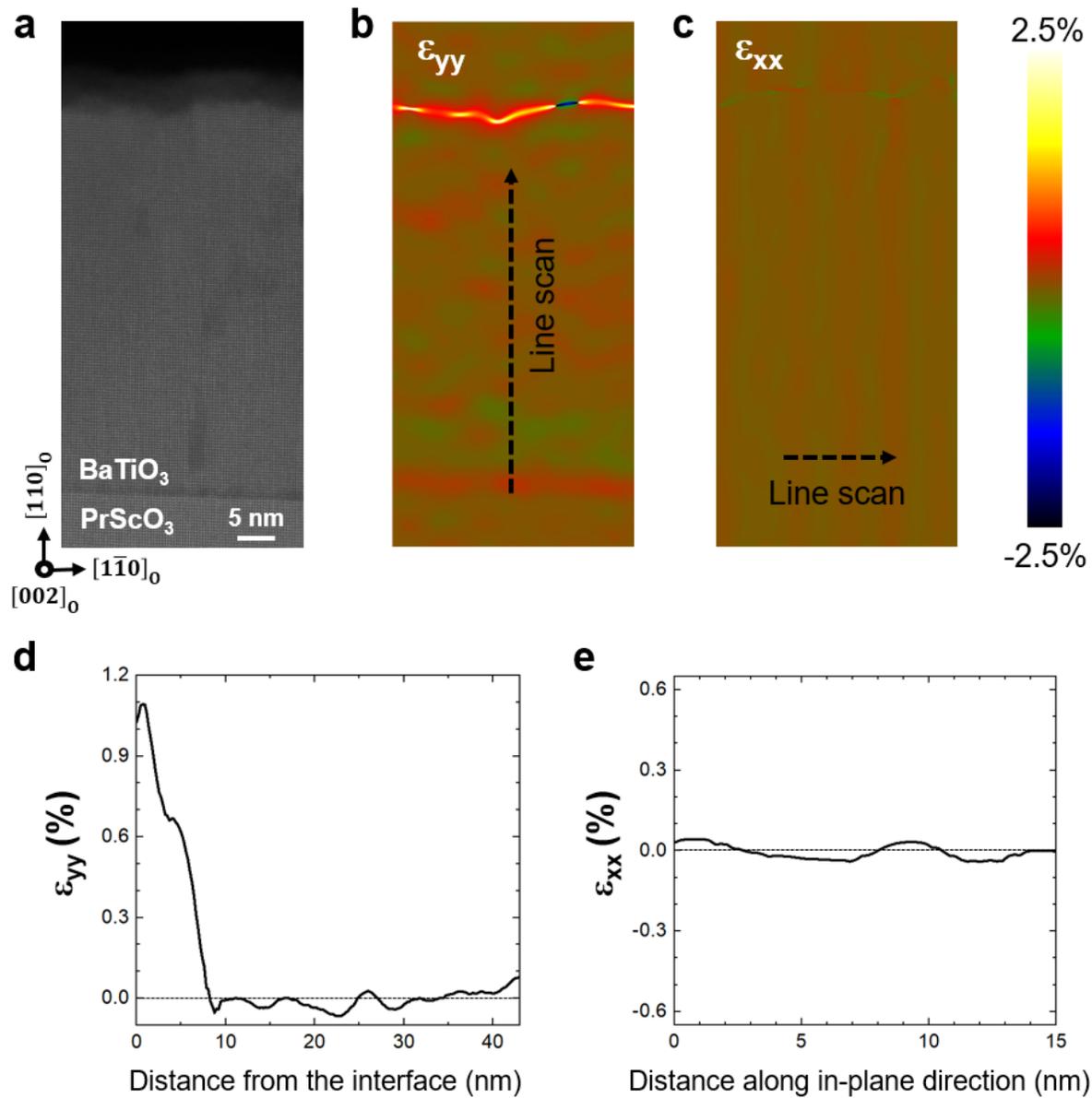

**Supplementary Figure 19. Geometric phase analysis (GPA) of the BTO film. a**, Low magnification HAADF STEM image of the BTO thin film. **b,c**, Corresponding GPA analysis along (b) y direction (out-of-plane) and (c) x direction (in-plane), respectively. The striped patterns in (c) are caused by scanning noise in the STEM image. **d,e**, Line scanning result of (d) $\varepsilon_{yy}$ as a function of distance from the interface, and (e) $\varepsilon_{yy}$ as a function of distance along in-plane direction.

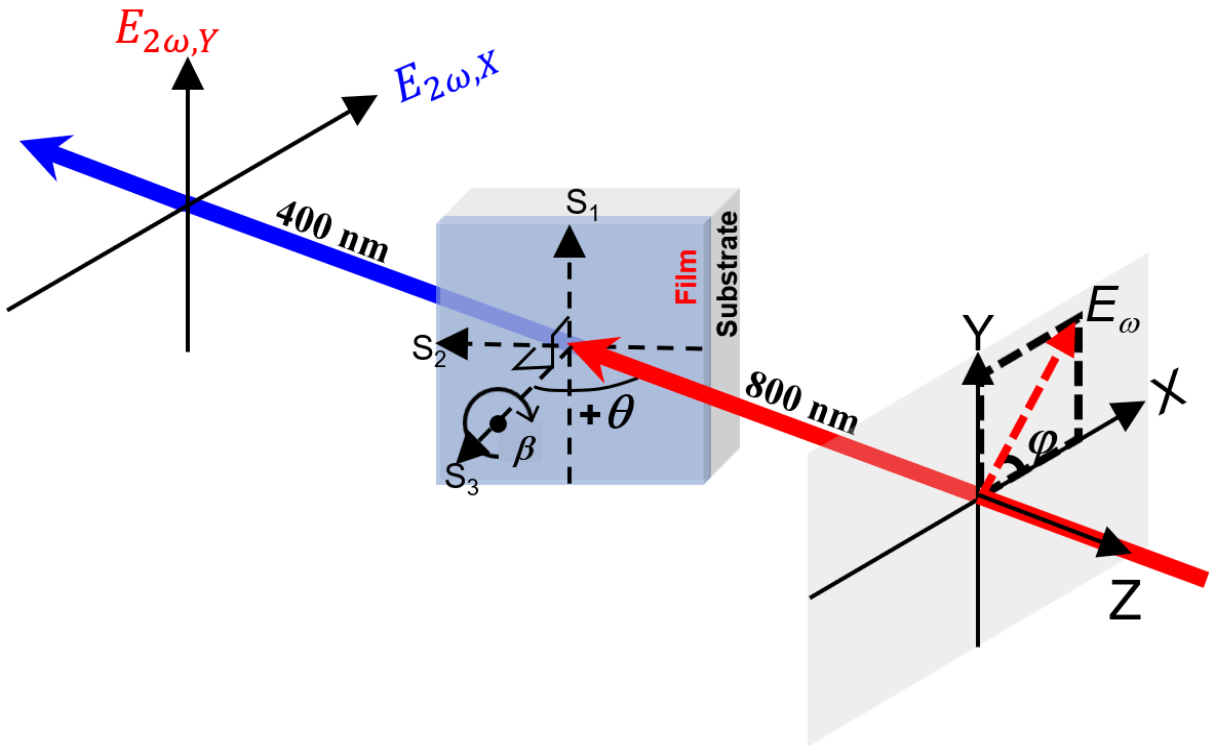

**Supplementary Figure 20. Schematic of far-field optical second harmonic generation setup**. Linear polarized fundamental optical beam at λ = 800 nm is incident onto the sample at an angle θ. Transmitted X-/Y-polarized SHG signal at λ = 400 nm is measured as rotating the polarizing direction φ.

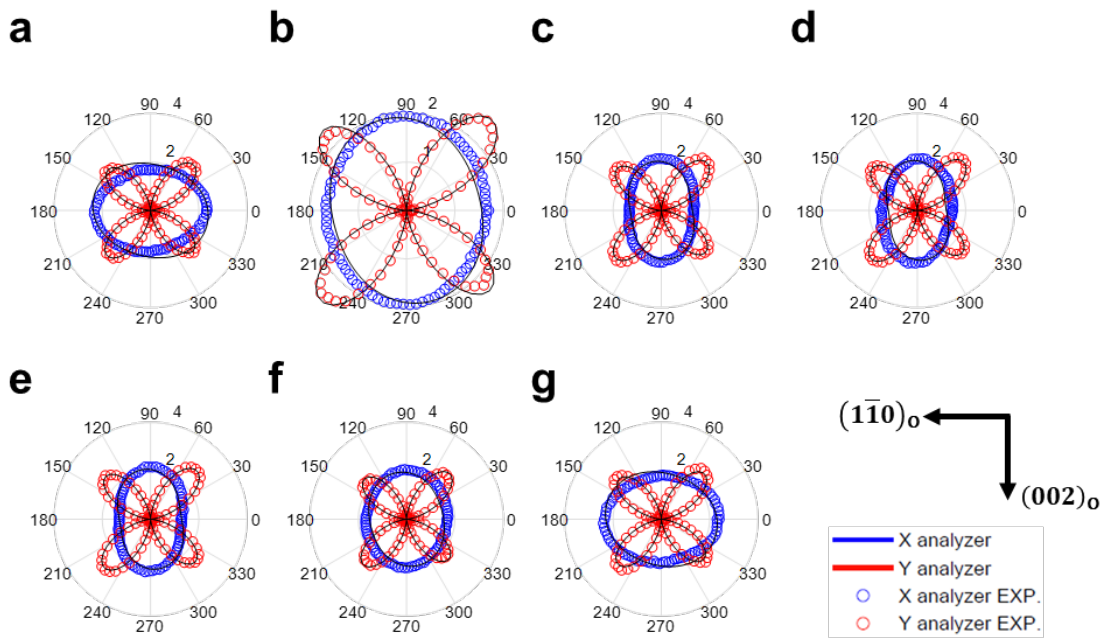

**Supplementary Figure 21. SHG polar plots of the BTO film on PSO $(110)_o$ substrate**. **a-g**, plots with an incident angle of -45° (**a**), -30° (**b**), -15° (**c**), 0° (**d**), 15° (**e**), 30° (**f**), and 45° (**g**). The solid lines are the fits assuming a monoclinic symmetry for BTO.

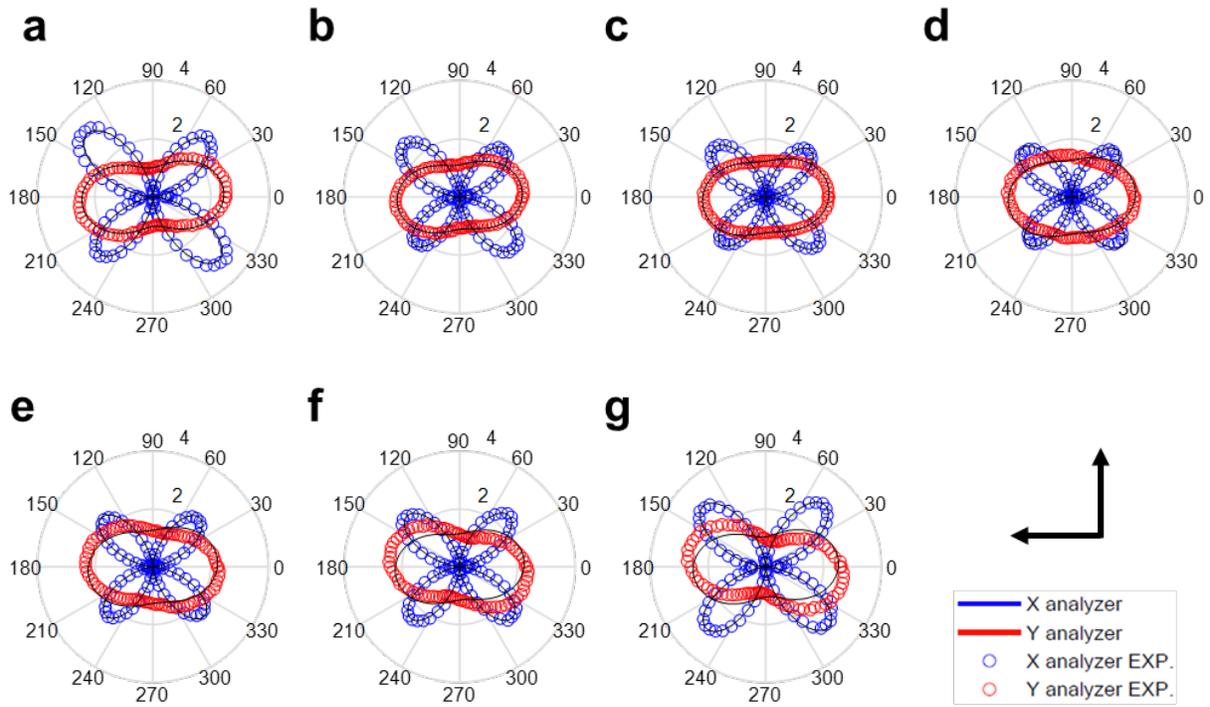

**Supplementary Figure 22. SHG polar plots of the BTO film on PSO $(110)_O$ substrate with different in-plane orientation**. **a-g**, plots with an incident angle of -45° (**a**), -30° (**b**), -15° (**c**), 0° (**d**), 15° (**e**), 30° (**f**), and 45° (**g**). Note that the in-plane sample orientation is rotated by 90° as compared to the data in Supplementary Figure 21. The solid lines are the fits assuming a monoclinic symmetry for BTO. The good quality of the fits as shown by the black solid lines in Supplementary Figure 21, 22 suggests the BTO film indeed exhibits a single ferroelastic domain with monoclinic distortion, which is consistent with STEM observations.

**Supplementary Table 1.** Anisotropic misfit strain between BaTiO$_3$ and REScO$_3$ (RE = La, Ce, Pr, Nd or Gd) substrates. The lattice parameters were referred from ref. 6, 7, except PSO. Pseudocubic lattice parameters of PSO were measured by XRD, using the single crystal substrate which was used in our experiment.

| Material | $a_{pc}$ (Å) | $b_{pc}$ (Å) | $c_{pc}$ (Å) | $c_{pc}/b_{pc}$ | $S_b$ (%) $(b_{RESO}-b_{BTO})/b_{BTO} \times 100$ | $S_c$ (%) $(c_{RESO}-c_{BTO})/c_{BTO} \times 100$ |
|---|---|---|---|---|---|---|
| BaTiO$_3$$^T$ | 3.992 | 3.992 | 4.036 | 1.011 | | |
| LaScO$_3$$^O$ | 4.053 | 4.049 | 4.053 | 1.001 | +1.43 | +0.42 |
| CeScO$_3$$^O$ | 4.036 | 4.023 | 4.036 | 1.003 | +0.78 | 0.00 |
| PrScO$_3$$^O$ | 4.026 | 4.007 | 4.026 | 1.005 | +0.38 | −0.25 |
| NdScO$_3$$^O$ | 4.013 | 4.000 | 4.013 | 1.003 | +0.20 | −0.57 |
| GdScO$_3$$^O$ | 3.973 | 3.967 | 3.973 | 1.002 | −0.63 | −1.56 |

pc: pseudocubic, T: tetragonal, O: orthorhombic

$a_{pc}//[110]_O$, $b_{pc}//[002]_O$, $c_{pc}//[1\bar{1}0]_O$

$S_b$: misfit strain in $b$-axis of BTO, $S_c$: misfit strain in $c$-axis of BTO